\documentclass[letterpaper,12pt,reqno]{amsart}

\usepackage{geometry}                 
\usepackage{hyperref}

\usepackage{booktabs,verbatim}
\usepackage{graphicx}

\usepackage{amsmath}
\usepackage{amssymb}
\usepackage{amsthm}
\usepackage{xcolor}

\usepackage{subfig}

\usepackage{slashed}

\usepackage[utf8]{inputenc}

\usepackage{amsfonts}

\usepackage{amsaddr}	

\usepackage{soul}

\usepackage{float}

\usepackage[backend=bibtex,style=numeric,citestyle=numeric,maxbibnames=99,giveninits,sorting=none,doi=false,isbn=false,url=false,eprint=false]{biblatex}
\renewbibmacro{in:}{\ifentrytype{article}{}{\printtext{\bibstring{in}\intitlepunct}}}	
\allowdisplaybreaks

\let\oldhat\hat
\renewcommand{\hat}[1]{\oldhat{\boldsymbol{#1}}}

\newcommand{\dee}{\ensuremath{\textrm{d}}}

\newcommand{\inty}[4]{\ensuremath{ \int_{#1}^{#2} \! #3 \, \dee#4 }}

\newcommand{\ip}[2]{\ensuremath{ \left< \left. #1 \right| #2 \right> } }

\newtheorem{definition}{Definition}		

\newtheorem{remark}{Remark}

\newtheorem{proposition}{Proposition}

\numberwithin{lemma}{section}
\numberwithin{example}{section}
\numberwithin{theorem}{section}
\numberwithin{corollary}{section}
\numberwithin{assumption}{section}
\newtheorem*{definition*}{Definition}		
\newtheorem*{assumption*}{Assumption}
\newtheorem*{remark*}{Remark}
\newtheorem*{theorem*}{Theorem}
\newtheorem*{lemma*}{Lemma}
\newtheorem*{proposition*}{Proposition}
\newtheorem*{corollary*}{Corollary}
\newtheorem*{example*}{Example}
\newtheorem*{conjecture*}{Conjecture}

\newcommand{\im}{\mathrm{i}}
\newcommand{\db}{{\rm d}}

\newcommand{\rA}{\mathrm{A}}
\newcommand{\rB}{\mathrm{B}}

\newcommand{\bsig}{\boldsymbol{\sigma}}

\newcommand{\Tr}{\mathrm{Tr}}

\definecolor{bvio}{rgb}{0.54, 0.17, 0.89}

\definecolor{fgr}{rgb}{0.13, 0.55, 0.13}

\newcommand{\nb}[3]{%
  {\colorbox{#2}{\bfseries\scriptsize\textcolor{white}{#1}}}%
  {\textcolor{#2}{\textit{#3}}}}
\newcommand{\dio}[1]{\nb{Dio}{bvio}{#1}}

\title[]{On the Su-Schrieffer-Heeger model of electron transport: Low-temperature optical conductivity by the Mellin transform}

\author{Dionisios Margetis}

\address[Dionisios Margetis]{Department of
      Mathematics, and Institute for Physical Science and Technology, University of Maryland, College Park, MD 20742, USA.}

\author{Alexander B. Watson, and Mitchell Luskin}

\address[Alexander B. Watson, Mitchell Luskin]{School of Mathematics,
    University of Minnesota, 
    Minneapolis, Minnesota 55455, USA.}
    
 \email{diom@umd.edu, watso860@umn.edu, luskin@umn.edu}     

\begin{document}

\begin{abstract}
	We describe the low-temperature optical conductivity as a function of frequency for a quantum-mechanical system of electrons that hop along a polymer chain. To this end, we invoke the Su-Schrieffer-Heeger \emph{tight-binding} Hamiltonian for non-interacting spinless electrons on a one-dimensional (1D) lattice. Our goal is to show via asymptotics how the interband conductivity of this system behaves as the smallest energy bandgap tends to close. \color{black} Our analytical approach includes: (i) the Kubo-type formulation for the optical conductivity with a nonzero damping due to microscopic collisions; \color{black} (ii) reduction of this formulation to a 1D momentum integral over the Brillouin zone; and (iii) evaluation of this integral in terms of elementary functions via the three-dimensional Mellin transform with respect to key physical parameters and subsequent inversion in a region of the respective complex space. Our approach reveals an intimate connection of the behavior of the conductivity to particular singularities of its Mellin transform. \color{black} The analytical results are found in good agreement with direct numerical computations.

	
\end{abstract}

\maketitle

\section{Introduction}
\label{sec:Intro}
The past several decades have seen significant advances in condensed matter physics, particularly the design, synthesis, modeling and applications of low-dimensional materials with unusual yet  practically appealing properties. These systems include conducting polymer chains~\cite{Heegeretal1988}, one-dimensional (1D)
nanowires~\cite{Ando2013}, and two-dimensional (2D) topological insulators with intriguing edge states~\cite{HasanKane2010}. We should also mention the celebrated graphene, a 2D semimetal, along with families of its variants such as 2D van der Waals heterostructures~\cite{Torres2014,CastroNetoetal2009,Geimetal2013,Lowetal2017,Peres2019}. Some of these materials offer novel paradigms of electronic transport~\cite{Peres2019,Falkovsky2007,LucasFong2018,Gurzhi1968}. When the mean free paths of electron-electron, electron-impurity and electron-phonon collisions are sufficiently small, 2D conducting systems may host nanoscale electromagnetic waves that challenge the classical diffraction limit~\cite{Lowetal2017}. From a quantum-mechanical view, this property is  related to features of  Hamiltonians for the wave motion of low-energy electrons in the underlying crystal (Bravais) lattices~\cite{CastroNetoetal2009,FeffWein2012,HasanKane2010}.

These developments give rise to the following broad question: How does the microscale motion of electrons in 1D and 2D materials, including topological insulators, affect the material optical response?
This question is not new; but its placement into the context of recent technological advances inspires mathematical problems that had eluded attention.

The theory of energy bands traditionally addresses the response to light of solids with periodic atomic potentials~\cite{AshcroftMermin-book,Kittel-book,Dresselhaus-notes}. The electromagnetic field is treated as classical, and is perturbatively coupled with Hamiltonians of low-energy non-interacting electrons~\cite{Mahan-book,Rickayzen-book}. This theory aims to explain in a simple fashion why  crystalline materials can be electric conductors or insulators. A key quantity is the optical conductivity $\bsig(\omega)$, a matrix-valued function of the frequency $\omega$. This $\bsig$ is macroscopically defined as a Fourier component of the coefficient entering the linear relation between the induced electric current density and the applied electric field. The microscopic origin of $\bsig$ was studied extensively; see the review article by Allen~\cite{Allen2006}. For example,  Kubo~\cite{Kubo1957} and Bellissard~\cite{Bellissard2002} make use of the trace of an operator involving current-current correlations. Usually, albeit not always~\cite{Cances2017}, the losses due to electron scattering  are modeled phenomenologically through a constant relaxation time, $\tau$~\cite{Bellissard1998,Bellissard2002,Peres2019,kubocomp20}. We adopt this view here.

In this paper, we carry out asymptotics to derive explicit formulas for the interband part, $\bsig^I(\omega)$, of the conductivity $\bsig(\omega)$ of a \emph{prototypical} 1D system, an electron hopping along a  polymer chain, in the zero-temperature limit. \color{black} The function $\bsig^I(\omega)$ is composed of contributions from matrix elements of the electron current operator that connect quantum states with distinct energies; these contributions lead to resonances of the conductivity at nonzero $\omega$. In 1D, $\bsig^I(\omega)$ reduces to a scalar function, $\sigma^I(\omega)$. \color{black} We compute this $\sigma^I(\omega)$ analytically by applying the three-dimensional (3D) Mellin transform to an integral for $\sigma^I(\omega)$ with respect to physical parameters. We show how $\sigma^I(\omega)$ is affected as the smallest energy bandgap, $\varepsilon_g$, tends to close. We believe that a novelty of our approach lies in the use of this multidimensional transform. \color{black}

We employ the Su-Schrieffer-Heeger (SSH) tight-binding model~\cite{SSH1979,Asboth-book}, which is a limit of the dynamics of a Schr\"odinger particle~\cite{https://doi.org/10.48550/arxiv.2107.09146}. \color{black} The model considered here is discrete in the configuration space and continuous in the momentum space; and has two energy bands at every momentum variable.

In physics it is generally known that, for fixed material parameters, $\bsig^I(\omega)$ has branch points at $\omega=\omega_i^R$, which correspond to energy band differences at critical points in the momentum space~\cite{Kittel-book}; the index $i$ counts the points $\omega_i^R$. The type of branch point depends on the spatial dimensionality, and other considerations. For typical textbook cases, see, e.g., Table 4.1, p.~35 in~\cite{Dresselhaus-notes}. By using the Mellin transform, we are able to analytically show how this behavior can be affected when a material parameter, particularly $\varepsilon_g$, is relatively small. We also show how the branch points of $\sigma^I(\omega)$ are intimately connected to singularities of the Mellin transform. To our knowledge, these aspects were previously unexplored.

We take into account the relatively small bandgap $\varepsilon_g$ and  relaxation rate $\tau^{-1}$ while the frequency $\omega$ varies in a reasonably wide range. Our study emphasizes distinct limiting procedures that come into play if the real frequency $\omega$ is close to any  ``resonance,'' i.e., if $|\omega-\omega_i^R|$, $\tau^{-1}$ and $\epsilon_g$ are simultaneously small, at low enough temperatures. We numerically demonstrate that our results are practically uniform in the frequency $\omega$. \color{black}

Our procedure can be outlined as follows. First, for the SSH model we derive a momentum integral for $\sigma^I$ over the Brillouin zone at nonzero temperatures. Then, we exactly evaluate the 3D Mellin transform of this integral with respect to physically appealing parameters. The transformed conductivity involves the Riemann zeta function and the Gamma function with arguments depending on linear combinations of (dual) complex variables. By inversion of this transform in a region of a complex space, we obtain $\sigma^I$ in the limit of zero temperature from a singularity of the integrand. We are unaware of similar applications of the multidimensional Mellin transform. \color{black}

In our analysis we relax mathematical rigor but provide an estimate for an error term germane to our low-temperature approximation for $\sigma^I(\omega)$. \color{black} We repeat that we numerically demonstrate the agreement of our asymptotics with direct numerical computations of the  momentum integral.

 {\em Notation.} Calligraphic capital letters, e.g., $\mathcal H$,  denote operators on a Hilbert space; but the ``density matrix'' is $\varrho$. The tilde on top of a symbol, e.g., $\widetilde I(\nu)$, denotes the Mellin transform of the respective function, e.g., $I(\epsilon)$. $f = O(g)$ ($f = o(g)$) means that $|f/g|$ is bounded by a nonzero constant (tends to  zero) in a prescribed limit. $f\sim g$ means $f-g=o(g)$. ``Schr\"odinger dynamics'' and ``Schr\"odinger particle'' imply the system evolution by Hamiltonians of the form $-\frac{\hbar^2}{2m}\Delta+V$; $m$ is the mass and $\hbar$ is the reduced Planck constant. \color{black} We use the $e^{-\im \omega t}$ single-frequency time dependence ($\im^2=-1$).

\subsection{Linear optical response, SSH model and problem statements}
\label{subsec:prob-stat}
Let us recall the abstract framework of linear response theory. The optical conductivity matrix, $\bsig$, is given by~\cite{Kubo1957,Bellissard1998,Bellissard2002}
\begin{subequations}\label{eqs:trace-cond-intro}
\begin{equation}\label{eq:cond_trace}
\bsig(\omega)=-4\sigma_0\,\Tr\left\{({\nabla}\mathcal H) \ \left(\mathcal L_{\mathcal H}-\im \hbar(\omega+ \im\tau^{-1})\right )^{-1}   ({\nabla} f(\mathcal H-\mu))\right\}~.
\end{equation}
In the above, $\mathcal H$ is the unperturbed electronic Hamiltonian, which acts on an appropriate Hilbert space $\mathfrak H$ ($\mathcal H:\,\mathfrak H\to \mathfrak H$); and, in the spirit of~\cite{Bellissard1998}, ${\nabla}\mathcal A:\,\mathfrak H\to \mathfrak H$ denotes the commutator $-\im\, [\mathcal X, \mathcal A]=-\im ( \mathcal X \mathcal A-\mathcal A \mathcal X)$ for any suitable $\mathcal A:\,\mathfrak H\rightarrow \mathfrak H$. Note that $f(\mathcal H)$ is the Fermi-Dirac distribution, viz.,
\begin{equation}\label{eq:F-D}
	f(\mathcal H)=\left(e^{\beta \mathcal H}+1 \right)^{-1}
\end{equation}
    where $\beta=1/T$ is the inverse absolute temperature. The quantity $\mu$ is the chemical potential, a Lagrange multiplier fixed by the total number, $N$, of non-interacting electrons; $N=\Tr\{f(\mathcal H -\mu)\}$. \color{black} The symbol $\mathcal L_{\mathcal H}$ denotes the Liouville superoperator for $\mathcal H$, which acts on any operator $\mathcal A:\,\mathfrak H\to \mathfrak H$ via $\mathcal L_{\mathcal H}(\mathcal A)=\im [\mathcal H, \mathcal A]=\im (\mathcal H \mathcal A-\mathcal A\mathcal H)$. The constant $\sigma_0=e^2/(4\hbar)$ has units of conductance and $e$ is the absolute value of the electron charge ($e>0$). The origin of~\eqref{eq:cond_trace} is reviewed in Appendix~\ref{app:conductivity-form}. The zero-temperature (as $\beta\to \infty$) limit of~\eqref{eq:F-D} is $\mathbf 1_{\mathcal H <0}(\mathcal H)$, the indicator function of the set $\{\mathcal H <0\}$. In what follows, we set $\mu = 0$; see Section \ref{subsec:past}.

Trace~\eqref{eq:cond_trace} can be computed via any suitable basis set. The \emph{interband} part, $\bsig^I(\omega)$, comes from extraction of the Drude conductivity, $\bsig^D$, viz.,
\begin{equation}\label{eq:interband-cond}
\bsig^I(\omega)=\bsig(\omega)-\bsig^D(\omega)~;\ \bsig^D(\omega)=-\frac{4\im\hbar^{-1} \sigma_0}{\omega+ \im\tau^{-1} }\Tr\left\{({\nabla}\mathcal H)   ({\nabla} f(\mathcal H))\right\}.
\end{equation}
\end{subequations}
If the eigenvectors of $\mathcal H$ are employed for the trace, then $\bsig^D$ contains only diagonal matrix elements of ${\nabla}\mathcal H$ and ${\nabla} f(\mathcal H)$. Hence, $\bsig^D$ is composed only of the \emph{intraband} transitions, to be contrasted to $\bsig^I$.

\begin{figure}
  \centering
\includegraphics[scale=.75]{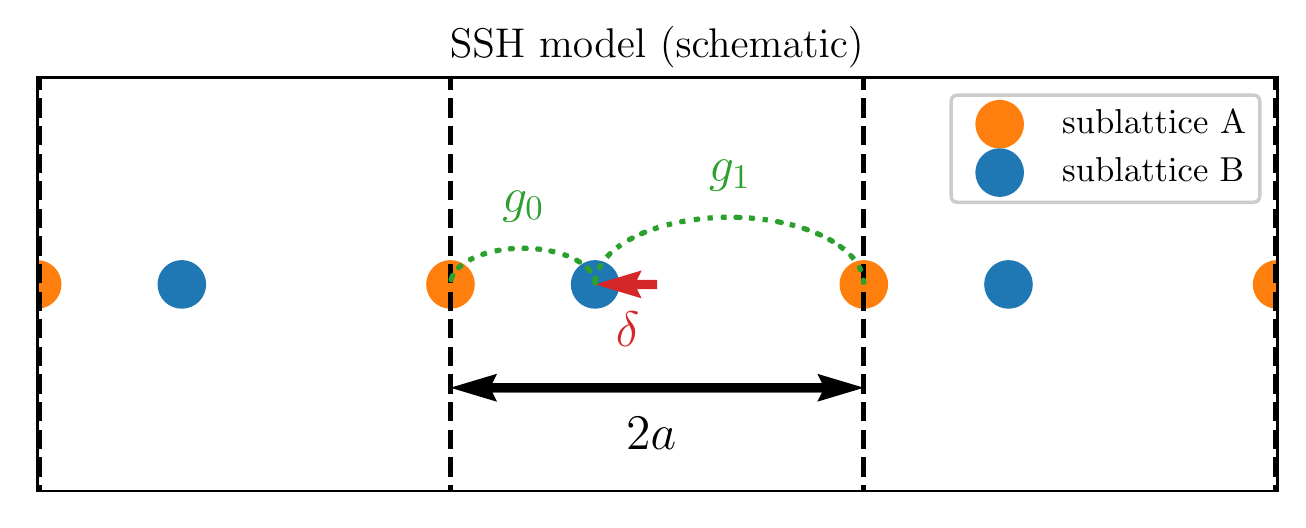}
  \caption{%
    Schematic of geometry and kinetics of the SSH model. Each fundamental cell has length $2a$; and the dimerization parameter $\delta$ is defined in Section~\ref{subsec:model}. Each cell contains A and B types of atoms (indicated by different colors). An electron hops between neighboring atoms of distinct types with rate $g_0$ within a cell and rate $g_1$ across cells.}
  \label{fig:SSH-kinetics}
\end{figure}

By the SSH model~\cite{SSH1979}, the electron Hilbert space is  $\mathfrak H=\ell^2(\mathbb{Z}; \mathbb{C}^2)$~\cite{Asboth-book}. Hence, $\mathfrak H$ is spanned by state vectors of the form $\psi_l^\alpha$ where $l\in \mathbb{Z}$ labels the lattice site and $\alpha\in \{\rA, \rB\}$ expresses the type of the atom per fundamental cell.  The tight-binding SSH Hamiltonian $\mathcal H$ is defined via the scheme~\cite{SSH1979,Asboth-book}
\begin{equation}\label{eq:SSH-scheme}
	(\mathcal H \psi)_l =
	\begin{pmatrix}
		-g_1 \psi_{l-1}^{\rB}-g_0 \psi_l^{\rB} \\
		-g_0 \psi_l^{\rA} - g_1 \psi_{l+1}^{\rA}
	\end{pmatrix}\quad \forall\, l\in\mathbb{Z}
\end{equation}
where $\psi_l = (\psi_l^{\rA},\, \psi_l^{\rB})^\top\in\mathbb{C}^2$; see Fig.~\ref{fig:SSH-kinetics}. \color{black} The constants $g_0,\, g_1$ are hopping rates. We assume $g_0\ge g_1> 0$, without loss of generality. In Section~\ref{subsec:model}, we review the connection of scheme~\eqref{eq:SSH-scheme} to the 1-particle Schr\"odinger  dynamics.

 \vskip5pt

\noindent {\bf Problem 1:} {\em By the formulation of~\eqref{eqs:trace-cond-intro} and~\eqref{eq:SSH-scheme}, derive a 1D integral in momentum space for the interband conductivity, $\sigma^I(\omega)$.}

Problem~1 calls for passing to the Bloch domain, which is natural since the system is translation invariant. We express the trace for $\sigma^I(\omega)$ in the eigenbasis of the Bloch-transformed Hamiltonian. \color{black} Let $I$ denote the requisite integral; see~\eqref{eqs:result-preview} in Section~\ref{subsec:results}.  We are unable to exactly compute $I$ in simple closed form in terms of known functions. Hence, we apply asymptotics.

\vskip5pt

\noindent {\bf Problem 2:} {\em Define the nondimensional parameters}
\begin{subequations}\label{eqs:SSH-pmts}
\begin{equation}\label{eq:eps-pmts-def}
	\epsilon_1:= \frac{(g_0-g_1)^2}{g_0 g_1}~,\ \epsilon_2:=\frac{4(g_0-g_1)^2-\hbar^2(\omega+\im\tau^{-1})^2}{4g_0 g_1}~,\ \epsilon_3:=\beta\sqrt{g_0 g_1}~.
\end{equation}
{\em Compute the integral $I$ for the interband conductivity to the leading order  in the low-temperature regime}
\begin{equation}\label{eq:eps-regime}
0< \epsilon_1\ll 1~,\quad \epsilon_3\sqrt{\epsilon_1}\gg 1~.
\end{equation}
\end{subequations}
%

The parameter $\sqrt{\epsilon_1}$ expresses the size of the smallest bandgap; $\epsilon_2$ depends on $\omega$, and signifies resonances; and $\epsilon_3$ measures the strength of the hopping energies relative to the absolute temperature, $T$. We repeat that the expressions for $\sigma^I(\omega)$ and integral $I$ are given in~\eqref{eqs:result-preview} (Section~\ref{subsec:results}). \color{black}

 The last condition in~\eqref{eq:eps-regime} is roughly suggested by the Fermi-Dirac distribution, $f(\mathcal H)$; cf.~\eqref{eqs:trace-cond-intro}. At low temperatures $e^{-\beta \mathcal H}$ should be small enough, where the exponent is controlled by $\beta$ (i.e., $\epsilon_3$) times the smallest energy scale, $\sqrt{\epsilon_1}$.  We will neglect such exponentially small terms. In Section~\ref{subsec:nonzero-T-rem}, we address the related error estimate for $\sigma^I$ by manipulation of the 1D momentum integral, and formally justify the parameter regime of~\eqref{eq:eps-regime}. \color{black}
%

%


\subsection{On the mathematical approach}
\label{subsec:math-approach}

For Problem~1, we employ the eigenvectors of the SSH Hamiltonian $\mathcal H$ along with~\eqref{eq:interband-cond}. Hence, we solve the eigenvalue problem for the spectrum of $\mathcal H$.

For Problem~2, we invert the (3D) Mellin transform of the requisite integral $I(\epsilon_1,\epsilon_2,\epsilon_3)$ for $\sigma^I(\omega)$. This technique can be powerful, although is not used often~\cite{Sasiela1993,Fikioris-book,AblowitzFokas-book,Carrier-book}; see Appendix~\ref{app:MT}. Applications of the 1D Mellin transform include, but are not limited to, the evaluation of  Feynman integrals~\cite{BjorkenWu1963,ChengWu-book}; and the computation of the radiated power of classical current distributions~\cite{MargetisFikioris2000}. We apply the 3D Mellin transform to an 1D integral for $\sigma^I$ in the zero-temperature limit. An ensuing task is to compute exactly certain power series from strings of poles in a dual variable, after the other two dual variables are approximately integrated out  (Section~\ref{subsec:zero-T}).  The extension of this transform technique to higher orders in the (temperature-dependent) parameter $e^{-\epsilon_3\sqrt{\epsilon_1}}$ presents difficulties that we leave unresolved; see Section~\ref{subsec:nonzero-T-rem}.

 The Mellin transform of $I(\epsilon_1,\epsilon_2,\epsilon_3)$ is defined by
\begin{subequations}
\begin{equation}\label{eq:I-MT}
\widetilde I(\lambda,\nu,\vartheta)=\int_0^\infty \int_0^\infty\int_0^\infty I(\epsilon_1, \epsilon_2,\epsilon_3)\,\epsilon_2^{-\nu}\epsilon_1^{-\lambda}\epsilon_3^{-\vartheta} \,\db\epsilon_3\,\db\epsilon_1\,\db\epsilon_2~.
\end{equation}
In the above, $(\Re\lambda, \Re\nu, \Re\vartheta)$ lies in some region $\mathbb{D}\subset\mathbb{R}^3$; and $\widetilde I$ is expressed in terms of the Gamma and Riemann zeta functions~\cite{Bateman-I}. See Proposition~\ref{prop:exact_MT}. We approximately invert $\widetilde I$ in the appropriate complex space, and write
\begin{equation}\label{eq:I-MT-inv}
I(\epsilon_1,\epsilon_2,\epsilon_3)=\frac{1}{(2\pi\im)^3}\int\limits_{\gamma_2-\im\infty}^{\gamma_2+\im \infty}
\int\limits_{\gamma_1-\im\infty}^{\gamma_1+\im \infty}
\int\limits_{\gamma_3-\im\infty}^{\gamma_3+\im \infty}\widetilde I(\lambda,\nu,\vartheta)\,\epsilon_2^{\nu-1}\epsilon_1^{\lambda-1}\epsilon_3^{\vartheta-1}\,\db\vartheta\,\db\lambda\,\db\nu~, 	
\end{equation}
\end{subequations}
where $(\gamma_1,\gamma_2,\gamma_3)\in \mathbb{D}$, by using a particular order of integrations. The zero-temperature limit of $I$ arises from a simple pole in the $\vartheta$-plane, for fixed $\lambda$ and $\nu$ (Section~\ref{subsec:zero-T}). The integration with respect to $\nu$, which is the variable dual to $\epsilon_2$, yields power series that are calculated in terms of hypergeometric functions. Our findings reduce to elementary functions capturing resonances. The main result is stated in Proposition~\ref{prop:zeroT-I}; and proved in Section~\ref{subsec:zero-T}.

\subsection{Physical motivation and assumptions}
\label{subsec:past}

The SSH Hamiltonian provides a quantum-mechanical toy model of electron transport in 1D. This model embodies some essential physics, while it is analytically tractable for the study of the interband conductivity as a function of frequency \emph{and} material parameters. \color{black}  Because of the idealizations involved, one may wonder if our results can offer insights into realistic situations. We invoke a minimal setting to analytically show how the behavior of the conductivity at singularities (branch points) in the $\omega$-plane is affected by microscale parameters. We believe that aspects of this behavior are universal, and must be described systematically. \color{black}

In general, the description of the optical response of a system when the parameters entering the unperturbed Hamiltonian take extreme values is practically compelling. This situation is relevant to physical systems in which a \emph{broken symmetry} of the Hamiltonian causes a small energy bandgap. Furthermore, in the celebrated multilayer graphene the optical conductivity can be altered through the associated kinetic rates, doping, and twist angle~\cite{Peres2019,CastroNetoetal2009,Falkovsky2007}. Our ultimate goal, which is not addressed here but partly motivates our work, is to apply a similar method to truly 2D materials~\cite{Haldane1988,CastroNetoetal2009,Peres2019}.

Besides the tight-binding character and dimensionality of the SSH model,  a few other simplifying assumptions should be spelled out. We \emph{phenomenologically} consider dissipative effects via a constant relaxation time, $\tau$~\cite{Allen2006}. We use a zero chemical potential $\mu$ ($\mu=0$) in the Fermi-Dirac distribution. This choice is consistent with the fact that  conducting polymers such as polyacetylene, usually described by the SSH model, have intriguing properties at low doping levels, i.e., near charge neutrality~\cite{Heegeretal1988}. \color{black} We also neglect couplings of the electronic Hamiltonian with lattice vibrations; thus, we assume that the hopping rates in the SSH model are (lattice-independent) constants~\cite{SSH1979}. \color{black}

\subsection{Article organization}
\label{subsec:organiz}

In Section~\ref{sec:SSH-results}, we review the SSH model and the tight-binding approach, and present two main results, Propositions~\ref{prop:exact_MT} and~\ref{prop:zeroT-I}. In Section~\ref{sec:asymptotics}, we prove Propositions~\ref{prop:exact_MT} and~\ref{prop:zeroT-I}, and provide a relevant low-temperature error estimate. \color{black} Section~\ref{sec:comp} focuses on comparisons of our analytical results to numerical computations of the requisite integral.
Section~\ref{sec:conclusion} concludes the paper.

\section{SSH model and main results}
\label{sec:SSH-results}
In this section, we review the SSH model and outline our results for the Mellin transform of the main integral and the asymptotic behavior of the interband conductivity, $\sigma^I$, as $T\to 0$. We also calculate the eigenvectors and spectrum of the Hamiltonian in the Bloch domain, needed in Section~\ref{sec:asymptotics}.

\subsection{SSH Hamiltonian: Definition and connection to Schr\"odinger particles}
\label{subsec:model}
We consider a 1D dimerized chain of atoms~\cite{SSH1979}. If the lattice constant is $2 a$, the Bravais lattice $\Lambda$ is defined by use of discrete position variable $R$ as
\begin{equation*}
    \Lambda := \left\{ R=2 a l\, :\, l \in \mathbb{Z} \right\}~.
\end{equation*}
A fundamental cell is $[0, 2a)$.
Within the $R$-th cell, there are two types of atoms (A and B) at positions $R + \tau^{\alpha}$ ($\alpha= \rA,\,\rB$) where $\tau^{\rA}= 0$ and $\tau^{\rB} = a + \delta$;
 $\delta \in (-a,a)$ is the dimerization parameter.

The electronic state vectors are modeled as elements $\psi$ of the Hilbert space $\mathfrak{H} := \ell^2(\mathbb{Z};\mathbb{C}^2)$. We denote such elements by $\left( \psi_{R} \right)_{R \in \Lambda} = \left( \psi^\rA_{{R}}, \psi^\rB_{{R}} \right)^\top_{{R} \in \Lambda}$ where $|\psi^\alpha_{{R}}|^2$ represents the electron density on sublattice $\alpha \in \{\rA,\rB\}$ in the ${R}$-th cell. We slightly modify the notation of Section~\ref{subsec:prob-stat}, replacing $l$ by $R$. The SSH Hamiltonian, $\mathcal H$, acts on $\psi$ according to~\eqref{eq:SSH-scheme}, which is now recast to
\begin{equation}\label{eq:SSH_H}
    \left(\mathcal H \psi \right)_R = \begin{pmatrix} - g_1 \psi_{R - 2 a}^{\rB} - g_0 \psi_{R}^{\rB} \\ - g_0 \psi_{R}^{\rA} - g_1 \psi_{R+2a}^{\rA} \end{pmatrix}~.
\end{equation}
Recall the schematic shown in Fig~\ref{fig:SSH-kinetics}. The parameters $g_0$ and $g_1$ have units of energy, and express hopping rates within the same cell or across neighboring cells, respectively. We take $g_0\ge g_1> 0$, without loss of generality.

\subsubsection{Connection to continuum Schr\"odinger dynamics}
\label{sssec:conn-Schroedinger}

We now sketch a \emph{formal} derivation of the model~\eqref{eq:SSH_H} from Schr\"odinger dynamics  (see, for example, Ashcroft and Mermin~\cite{AshcroftMermin-book}), although we will discuss shortly why this argument should be regarded with some skepticism. Let $V_{\text{at}}(x)$ denote the atomic potential, which is a real function such that the atomic continuum Schr\"odinger operator
    $- \frac{\hbar^2}{2m}\Delta + V_{\text{at}}(x)$
has a \emph{non-degenerate} ground state, $\Phi(x)$. Then, define the continuum Schr\"odinger operator (with $\hbar=1=2m$)
\begin{equation} \label{eq:SSH_cont}
    \mathcal H_{\text{cont}} := - \Delta + V(x), \quad V(x) := \sum_{R \in \Lambda} V_{\text{at}}(x - (R + \tau^A)) + V_{\text{at}}(x - (R + \tau^B))~.
\end{equation}
Model \eqref{eq:SSH_H} emerges when~\eqref{eq:SSH_cont} is projected onto the subspace of $L^2(\mathbb{R})$ generated by translations of the atomic ground state, $\{ \Phi(x - (R + \tau^A)), \Phi(x - (R + \tau^B)) \}_{R \in \Lambda}$, with neglect of matrix elements corresponding to interactions between atomic ground states that are separated beyond nearest neighbors. \color{black} This approximation is formally justified assuming sufficient decay of the wave function $\Phi$. The components $\psi^\alpha_R$ appearing in \eqref{eq:SSH_H} correspond to the coefficients of the translated ground states $\Phi(x - (R + \tau^\alpha))$ for every $R \in \Lambda$ and $\alpha \in \{A,B\}$, while the coefficients $g_0, g_1$ denote the overlap integrals between nearest-neighbor atomic potentials, i.e., $g_0 \approx \ip{\Phi(\cdot - \tau^A)}{\mathcal{H}_{\text{cont}} \Phi(\cdot - \tau^B)}_{L^2}$, $g_1 \approx \ip{\Phi(\cdot - \tau^B)}{\mathcal{H}_{\text{cont}} \Phi(\cdot + 2 a - \tau^A)}_{L^2}$; $\langle \cdot | \cdot\rangle_{L^2}$ is the $L^2$-inner product.

A \emph{rigorous} derivation of \eqref{eq:SSH_H} from \eqref{eq:SSH_cont} was carried out by Shapiro, Fefferman and Weinstein in a sequence of papers~\cite{https://doi.org/10.48550/arxiv.2006.08025,https://doi.org/10.48550/arxiv.2010.12097,https://doi.org/10.48550/arxiv.2107.09146}, following earlier work of Lee-Thorp, Fefferman and Weinstein~\cite{Fefferman2018} as well as work of Helffer and
Sj{\"o}strand~\cite{Helffer1984,10.1007/3-540-51783-9_19}. The basic idea is to replace $V$ by $\breve\lambda^2 V$ and then consider the limit of large $\breve\lambda$. (This is, equivalently, the semi-classical, or deep-well limit.) Two important subtleties arise in the derivation \cite{https://doi.org/10.48550/arxiv.2107.09146} which are worth emphasizing, since they indicate the limitations of the formal argument given previously. First, to rigorously derive \eqref{eq:SSH_H} with non-zero dimerization parameter $\delta \neq 0$, the distances between atoms in the model \eqref{eq:SSH_cont} must be \emph{scaled with $\breve \lambda$}. Second, the topological classification (in the sense of Kitaev's table of topological insulators \cite{doi:10.1063/1.3149495}) of the discrete SSH model emerges only in the tight-binding limit. In particular, topologically \emph{distinct} discrete SSH models \eqref{eq:SSH_H} may emerge from topologically \emph{equivalent} continuum SSH models \eqref{eq:SSH_cont}.

\subsubsection{Diagonalization of SSH Hamiltonian in Bloch domain}
\label{sssec:diagon-SSH}

The SSH Hamiltonian, $\mathcal H$, is invariant under lattice translations. Hence, it is natural to pass to the Bloch domain, $\mathfrak H_*$ (defined below). Accordingly, we introduce the reciprocal lattice constant $b$, and the reciprocal Bravais lattice
\begin{equation*}
    \Lambda_* := \left\{ G = b n : n \in \mathbb{Z} \right\}~; \quad b := \frac{\pi}{a}~.
\end{equation*}
We take $[0, b)$ as a fundamental cell of this lattice (Brillouin zone).

The Bloch domain is $\mathfrak H_*=L^2([0, b);\mathbb{C}^2)$. In this domain, the electronic state vectors are written as
\begin{equation*}
\oldhat{\psi} = \left( \oldhat{\psi}(k) \right)_{k \in [0,b)} = \left( \oldhat{\psi}^A(k), \oldhat{\psi}^B(k) \right)^\top_{{k} \in [0, b)}~.
\end{equation*}
The unitary Bloch transform $\mathcal{G}\, :\,  \mathfrak{H}\rightarrow \mathfrak{H}_*$ and its inverse are defined by
\begin{equation*}
    \begin{split}
        [ \mathcal{G} \psi ]^\alpha({k}) &:= \frac{1}{\sqrt{b}}\sum_{{R} \in \Lambda} e^{- \im {k} ({R} + {\tau}^\alpha)} \psi^\alpha_{{R}}=: \oldhat{\psi}^\alpha(k)~, \\
        \left[ \mathcal{G}^{-1} \oldhat{\psi}\right]^\alpha_{{R}} &:= \frac{1}{\sqrt{b}} \inty{0}{b}{ e^{\im {k} ({R} + {\tau}^\alpha)} \oldhat{\psi}^\alpha({k}) }{{k}}~, \color{black}  \quad \alpha \in \{\rA,\rB\}~.
    \end{split}
\end{equation*}

The system Hamiltonian is block diagonal on $\mathfrak{H}_*$, taking the form
\begin{subequations}\label{eqs:H-Bloch}
\begin{equation}\label{eq:H-Bloch-form}
    \left( \mathcal{G} \mathcal H \mathcal{G}^{-1} \oldhat{\psi} \right)({k}) =: \oldhat{\mathcal H}({k}) \oldhat{\psi}({k})~,
\end{equation}
where
\begin{equation} \label{eq:graphene_Bloch}
    \oldhat{\mathcal H}({k}) = - \begin{pmatrix} 0 & F({k}) \\ F^*({k}) & 0 \end{pmatrix}, \quad F({k}) = e^{\im {k} ( {\tau}^{\rB} - {\tau}^{\rA} )} ( g_0 + g_1 e^{- 2 \im  {k}a} )
\end{equation}
\end{subequations}
and $F^*$ denotes the complex conjugate of $F$.
The Hamiltonian $\oldhat{H}({k})$ is directly diagonalized. It has
eigenpairs
\begin{subequations}\label{eqs:monolayer_bands}
\begin{equation} \label{eq:monolayer_bands}
    \varepsilon_s({k}) = s |F({k})|, \quad \varphi_s({k}) = \frac{1}{\sqrt{2}} \left( 1 , -s e^{-\im\chi(k)} \right)^\top~;\quad  e^{\im \chi(k)}:=\frac{F({k})}{ | F({k}) | }
\end{equation}
and $s=\pm$. The functions $\varepsilon_\pm : [0, b) \rightarrow \mathbb{R}$ are the Bloch bands; these are
\begin{equation}\label{eq:monolayer_bands-en}
\varepsilon_\pm(k)=\pm\sqrt{(g_0-g_1)^2+4g_0g_1\cos^2(ka)},\qquad k\in [0, \pi/a)~.	
\end{equation}
\end{subequations}
For every $k$, $\varepsilon_+(k)-\varepsilon_-(k)$ is the bandgap. The smallest bandgap is $\varepsilon_g:=2|g_0-g_1|=2(g_0-g_1)$ and the largest one is $2(g_0+g_1)$. We often refer to the former as the ``small bandgap'' and the latter as the ``large bandgap.'' \color{black}

We should note that the dependence of the Bloch Hamiltonian \eqref{eq:graphene_Bloch} on $\delta$ through $\tau^\text{B} - \tau^\text{A}$ can always be removed by a gauge transformation; or, equivalently, by a redefinition of the Bloch transform. For simplicity of our presentation, we therefore set $\tau^\text{B} - \tau^\text{A} = a$ henceforth. \color{black}

\subsection{Results on 3D Mellin transform and zero-temperature asymptotics}
\label{subsec:results}
We show that by~\eqref{eq:interband-cond} the conductivity $\sigma^I(\omega)$ can be written as
\begin{subequations}\label{eqs:result-preview}
\begin{equation}\label{eq:sigmaI-prev}
\sigma^I(\omega)=\frac{\im a\sigma_0}{2} \hbar (\omega+\im\tau^{-1})(g_0^2-g_1^2)^2 (g_0g_1)^{-5/2} I(\epsilon_1, \epsilon_2, \epsilon_3)~,	
\end{equation}
where $I$ can be expressed as a 1D integral over the Brillouin zone (Section~\ref{subsec:integral}). By a change of variable, integral $I$ is recast to the contour integral
\begin{equation}\label{eq:I-int-def}
I(\epsilon_1,\epsilon_2, \epsilon_3):=\frac{1}{2\pi \im}\oint_{\{|z|=1\}}\frac{\breve f(z;\epsilon_1,\epsilon_3)-\breve f(z;\epsilon_1,-\epsilon_3)}{\{\epsilon_1+r(z)\}^{3/2}}\	\frac{1}{\epsilon_2+r(z)}\ \frac{\db z}{z}~.
\end{equation}
In the above, we define the following functions of the complex variable $z$:
\begin{equation}\label{eq:FD-z}
	\breve f(z; \epsilon_1,\epsilon_3):=\left(1+e^{ \epsilon_3\sqrt{\epsilon_1+r(z)}}\right)^{-1}~,\quad r(z):=\frac{(z+1)^2}{z}~;
\end{equation}
\end{subequations}
$\epsilon_j$ ($j=1,\,2,\,3$) are the parameters introduced in~\eqref{eq:eps-pmts-def}. Evidently, $\breve f(z;\epsilon_1,\epsilon_3)$ is the Fermi-Dirac distribution, $f(\varepsilon_+)$, at the energy band $\varepsilon_+(k)$ under the mapping $k\mapsto z$ with $z=e^{2\im ka}$. This transformation maps the Brillouin zone onto the unit circle in the $z$-plane.

By~\eqref{eq:I-MT} we evaluate the Mellin transform of $I(\epsilon_1,\epsilon_2, \epsilon_3)$ exactly; see Section~\ref{subsec:MellinT} for details. The result can be stated as follows.
\begin{proposition}
The Mellin transform of integral $I(\epsilon_1, \epsilon_2, \epsilon_3)$ equals
\begin{align}\label{eq:I-MT-exact}
\widetilde I(\lambda,\nu,\vartheta)&=-2^{-2\nu-2\lambda +\vartheta-2}\pi^{\frac12-\vartheta}\frac{\zeta\left(\vartheta,\frac12\right)}{\sin\left(\pi\vartheta/2\right)}\notag\\
&\times \frac{\Gamma(1-\lambda)\,\Gamma(\lambda-\vartheta+\frac32)\,\Gamma(1-\nu)\,\Gamma(\nu)\,\Gamma(-\nu-\lambda+\vartheta-1)}{\Gamma(-\frac12-\lambda-\nu+\vartheta) \,\Gamma(\frac52-\vartheta)}~;
\end{align}
$\zeta(\vartheta,\varsigma)$ is the generalized zeta function. Integral~\eqref{eq:I-MT} converges in the region
\begin{equation*}
	\mathbb{D}=\left\{(\gamma_1, \gamma_2, \gamma_3): -\frac12<\gamma_1<1,\, 0<\gamma_2 <\frac12,\, 1<\gamma_3< 2,\, 1+\gamma_2<\gamma_3 -\gamma_1 <\frac32    \right\}
\end{equation*}
where $(\gamma_1, \gamma_2, \gamma_3)=(\Re\lambda, \Re\nu, \Re\vartheta)$.
 \label{prop:exact_MT}
\end{proposition}

The Mellin transform is reviewed in Appendix~\ref{app:MT}. We show that
\begin{equation}\label{eq:gen-zeta-R}
\zeta\bigl(\vartheta,\textstyle{\frac12}\bigr)=(2^\vartheta-1)\zeta(\vartheta)
\end{equation}
where $\zeta(\vartheta)$ is the Riemann zeta function~\cite{Bateman-I}; see Appendix~\ref{app:zeta}. Our definitions of the parameters $\epsilon_j$ ($j=1,\,2,\,3$) are crucial for obtaining the result of Proposition~\ref{prop:exact_MT}. Furthermore, we are able to extract a simple asymptotic formula for $\sigma^I(\omega)$ near the small-bandgap resonance, as outlined below. \color{black}

By~\eqref{eq:I-MT-inv} we invert transform~\eqref{eq:I-MT-exact} in the parameter regime of~\eqref{eq:eps-regime}. The main result is stated as follows:
\begin{proposition}[Zero-temperature limit of interband conductivity]   Let us assume that  $0< \epsilon_1\ll 1$ and $\epsilon_3\sqrt{\epsilon_1} \gg 1$. \color{black} Then, the integral $I(\epsilon_1,\epsilon_2,\epsilon_3)$ entering~\eqref{eq:sigmaI-prev} and~\eqref{eq:I-int-def} is expressed by the asymptotic formula
\begin{align}\label{eq:I-asympt-zeroT}
	I&\sim -\frac{1}{\pi}\frac{1}{\epsilon_1\epsilon_2}\left(1-\frac{\epsilon_1}{\epsilon_2}\right)^{-1}\left\{1-\frac{\epsilon_1}{\epsilon_2}\left(1-\frac{\epsilon_1}{\epsilon_2}\right)^{-\frac12}\ln\Biggl(\sqrt{\frac{\epsilon_2}{\epsilon_1}}+\sqrt{\frac{\epsilon_2}{\epsilon_1}-1}\Biggr)\right\}  \notag \\
	& -\frac{1}{16\pi}\frac{1}{\epsilon_2}\left\{-1+\frac{64}{\epsilon_2}\,\frac{\sin^{-1}\bigl(\sqrt{-4/\epsilon_2}\bigr)}{\sqrt{-4/\epsilon_2}}\frac{1}{\sqrt{1+4/\epsilon_2}}-\frac{8}{\epsilon_2}\ln\left(\frac{16}{\epsilon_2}\right)    \right\}~.
\end{align}
\label{prop:zeroT-I}
\end{proposition}

\begin{remark}\label{rmk:2reson}
{\rm Formula~\eqref{eq:I-asympt-zeroT} is obtained from~\eqref{eq:sigmaI-prev} by taking the limit as $\beta\to \infty$ (or, $T\to 0$); and then expanding in $\epsilon_1$. This result captures both the small- and the large-bandgap resonances. For fixed $\epsilon_1$ and $\tau\to \infty$, in the former type of resonance we have $\omega\to  \pm \epsilon_g=\pm 2(g_0-g_1)$, or $\epsilon_2\to 0$; while in the latter type of resonance we have $\omega\to \pm 2(g_0+g_1)$, or $1+\frac{4}{\epsilon_2}\to 0$.	}
\end{remark}

\begin{remark}\label{rmk:sigmaI-behavior}
{\rm For real $\omega$, we now compare predictions from~\eqref{eq:I-asympt-zeroT} \emph{near the small-bandgap resonance} to the typical textbook case for the behavior of $\sigma^I(\omega)$ near a resonance~\cite{Dresselhaus-notes}, in view of our condition $\epsilon_1\ll 1$. We distinguish two limiting procedures involving the ratio $\epsilon_1/\epsilon_2$, which is controlled by $|\omega\mp 2(g_0-g_1)|/\left(2(g_0-g_1)\right)$ and $(g_0-g_1)\tau$ near this resonance, where $\omega\sim \pm 2(g_0-g_1)$.
To start with, let $(g_0-g_1)\tau\to \infty$ while $|\omega^2-4(g_0-g_1)^2|/\left(4(g_0-g_1)^2\right)$ is fixed and small. It can be shown that this case amounts to taking $\epsilon_2\to 0$ while $\epsilon_1$ is fixed and small (thus, $\epsilon_2/\epsilon_1\to 0$). Expanding~\eqref{eq:I-asympt-zeroT} in powers of $\epsilon_2/\epsilon_1$, we find $I\sim -(\im/\pi) \epsilon_1^{-3/2} \epsilon_2^{-1/2}$. By~\eqref{eq:sigmaI-prev}, we obtain}
\begin{subequations}\label{eqs:sigmaI-reson}
\begin{align}
	\frac{\sigma^I(\omega)}{2\sigma_0 a}\sim \frac{1}{2\pi}\frac{(g_0+g_1)^2}{g_0 g_1}\frac{1}{\sqrt{\epsilon_2}}\sim  \frac{1}{\pi}\frac{(g_0+g_1)^2}{\sqrt{g_0 g_1}}\frac{1}{\sqrt{4(g_0-g_1)^2-\omega^2-2\im \omega \tau^{-1}}}~, \label{eq:sigmaI-reson1}
\end{align}	
{\rm which reduces to the typical textbook case in 1D~\cite{Dresselhaus-notes} if $2\im \omega \tau^{-1}$ is neglected. Formula~\eqref{eq:sigmaI-reson1} holds if $|\epsilon_2|\ll \epsilon_1\ll 1$, or $4(g_0-g_1)\tau^{-1}\lesssim
 |\omega^2-4(g_0-g_1)^2|\ll 4(g_0-g_1)^2$, near the small-bandgap resonance. Now consider a different limiting procedure near this resonance: Let $(g_0-g_1)\tau\to 0$ while $|\omega^2-4(g_0-g_1)^2|/\left(4(g_0-g_1)\tau^{-1}\right)$ is small; thus, take $\epsilon_1\to 0$ while $\epsilon_2$ is fixed and possibly small (thus, $\epsilon_1/\epsilon_2\to 0$). By expanding~\eqref{eq:I-asympt-zeroT} in powers of $\epsilon_1/\epsilon_2$, we obtain $I\sim -(1/\pi) \epsilon_1^{-1} \epsilon_2^{-1}$. Hence, by~\eqref{eq:sigmaI-prev} we find}
    \begin{align}
	\frac{\sigma^I(\omega)}{2\sigma_0 a}&\sim \frac{1}{2\pi}\frac{(g_0+g_1)^2}{g_0 g_1}\frac{1}{\sqrt{\epsilon_2}} \sim \frac{1}{\pi}\frac{(g_0+g_1)^2}{\sqrt{g_0 g_1}}\frac{1}{\sqrt{\tau^{-2}-2\im\omega\tau^{-1}}}~, \label{eq:sigmaI-reson2}
\end{align}
\end{subequations}
   {\rm  which becomes linear with $\tau$ if $2\im \omega\tau^{-1}$ is neglected. At $\omega = 0$, \eqref{eq:sigmaI-reson2} resembles the intraband contribution to the conductivity, which is consistent with the reduction of the SSH model to a one-band model when $\epsilon_1 \rightarrow 0$. Approximation~\eqref{eq:sigmaI-reson2} is reasonable if $\epsilon_1\ll |\epsilon_2|\ll 1$, or $|\omega^2-4(g_0-g_1)^2|\ll 4(g_0-g_1)\tau^{-1}$ and $4(g_0-g_1)^2\ll 4(g_0-g_1)\tau^{-1}$ while $\sqrt{g_0 g_1}\tau\gg 1$. Note that $\sigma^I(\omega)$ has the same asymptotic form as a function of $\epsilon_2$ in the two limiting cases; cf.~\eqref{eq:sigmaI-reson1}.}
\end{remark}
\begin{remark}\label{rmk:complex-omega}
{\rm We may extend the above study to \emph{complex} frequencies $\omega$ by allowing $\omega+\im \tau^{-1} \sim  \pm 2(g_0-g_1)$, when $\tau$ is finite and nonzero. In the limits $\omega+\im \tau^{-1} \to  \pm 2(g_0-g_1)$, $\sigma^I(\omega)$ exhibits (two) branch points in the $\omega$-plane that correspond to the small bandgap. The behavior of $\sigma^I(\omega)$ in the vicinity of each branch point is sensitive to the ratio $\epsilon_1/\epsilon_2$. Indeed, in this vein we can show that $\frac{\sigma^I(\omega)}{2\sigma_0 a}=O(|\epsilon_2|^{-1/2})$ as $\epsilon_1\to 0$ with $\epsilon_2/\epsilon_1\to 0$; while $\frac{\sigma^I(\omega)}{2\sigma_0 a}=O(\sqrt{\epsilon_1}/|\epsilon_2|)$ when $\epsilon_2\to 0$ with $\epsilon_1/\epsilon_2\to 0$. An underlying property is that $\frac{\omega+\im \tau^{-1}}{\sqrt{g_0 g_1}}=O(\sqrt{\epsilon_1})$ regardless of the order of magnitude of $\tau$. Compare to the case with real $\omega$  (Remark~\ref{rmk:sigmaI-behavior}). A branch point of the same type occurs if $\epsilon_2\to -4$, for the large bandgap; but the behavior of $\sigma^I(\omega)$ in the vicinity of this point is \emph{not} affected by $\epsilon_1$. This is expected  because the band structure of the system near this resonance is insensitive to $\epsilon_1$.}
\end{remark}
\begin{remark}\label{rmk:on-proof-Props12}
{\rm For our proofs of Propositions~\ref{prop:exact_MT} and~\ref{prop:zeroT-I}, see Sections~\ref{subsec:MellinT} and~\ref{subsec:zero-T}, respectively. In the proof of Proposition~\ref{prop:zeroT-I} (Section~\ref{subsec:zero-T}) we focus on the derivation of formula~\eqref{eq:I-asympt-zeroT} directly from the exact 3D Mellin transform of $I(\epsilon_1, \epsilon_2, \epsilon_3)$. The role of the condition $\epsilon_3\sqrt{\epsilon_1}\gg 1$ is discussed in Section~\ref{subsec:nonzero-T-rem}. }
 \end{remark} \color{black}

Numerical computations for comparison purposes are carried out in Section~\ref{sec:comp}.
In the remainder of the paper, we set $\hbar=1$ for ease of notation.  \color{black}

\section{Asymptotic evaluation of $\sigma^I(\omega)$ by the Mellin transform}
\label{sec:asymptotics}
In this section, we derive a 1D integral representation for $\sigma^I(\omega)$ within the SSH model. Furthermore, we prove Propositions~\ref{prop:exact_MT} and~\ref{prop:zeroT-I}, and discuss the nature of a possible correction term in the low-temperature expansion.

First, let us generally discuss the computation of trace~\eqref{eq:cond_trace} in terms of matrix elements in a convenient basis. Extending the notation of Section~\ref{subsec:model} to $d$ spatial dimensions, we employ the eigenbasis $\{\oldhat{\varphi}_s(k)\}_s$, which consists of the eigenvectors of the unperturbed Hamiltonian in the Bloch domain, $\mathfrak H_*$. The index $s$ ($s=1,\,\ldots,\,n_b$) counts energy bands, $k$ is in the Brillouin zone, and $\oldhat{\varphi}_s(k)\in \mathbb{C}^{n_b}$.  By use of~\eqref{eq:interband-cond}, the integral for $\bsig^I$ is ($l, m=1,\,\ldots,\,d$)~\cite{Peres2019}
\begin{equation}\label{eq:interb-cond-matrix}
	\sigma_{lm}^I=-\frac{\im 4\sigma_0}{(2\pi)^d}\sum_{s\neq s'}\inty{\mathrm{BZ}}{}{\frac{\langle s | \partial_{k_l}\oldhat{\mathcal H}|s'\rangle \langle s' | \partial_{k_m}\oldhat{\mathcal H}|s\rangle}{ \varepsilon_{ss'}(k)+\omega+\im \tau^{-1}} \frac{f(\varepsilon_s(k))-f(\varepsilon_{s'}(k))}{\varepsilon_{ss'}(k)}}{k}
\end{equation}
where the integration is carried out over the Brillouin zone, denoted as BZ.
In the above, $\oldhat{\mathcal H}(k)$ is the Bloch-transformed Hamiltonian (an $n_b\times n_b$ matrix), $\langle s | \partial_{k_l}\oldhat{\mathcal H}|s'\rangle:=\oldhat{\varphi}_s^*(k)^\top (\partial_{k_l}\oldhat{\mathcal H}) \oldhat{\varphi}_s(k)$,  $\varepsilon_{ss'}(k):=\varepsilon_s(k)-\varepsilon_{s'}(k)$, $\varepsilon_s(k)$ is the $s$-th eigenvalue (band) of $\oldhat{\mathcal H}(k)$, and $k_l$ is the $l$-th component of momentum $k$.

In particular, for the SSH model (Section~\ref{subsec:model}), we have $d=1$ and $n_b=2$. Thus, let $s, s'=\pm$. The energy bands are $\varepsilon_\pm(k)$ ($\varepsilon_+=-\varepsilon_->0$).  The matrix $[\sigma_{lm}]$ ($[\sigma_{lm}^I]$) reduces to  a scalar, $\sigma$ ($\sigma^I$).

\subsection{SSH model: Integral for $\sigma^I(\omega)$ over Brillouin zone}
\label{subsec:integral}
Consider the unperturbed Hamiltonian~\eqref{eq:graphene_Bloch}, in the Bloch domain. In view of formula~\eqref{eq:interb-cond-matrix}, we need to compute matrix elements of $\partial_k\oldhat{\mathcal H}$ in the eigenbasis $\{\oldhat{\varphi}_s(k)\}_s$. A direct calculation using the eigenpairs of~\eqref{eqs:monolayer_bands} yields
\begin{align*}
	\langle s|\partial_k \oldhat{\mathcal H}|s' \rangle &=-\frac{1}{2}\left(1,\,-s e^{\im \chi(k)}\right)
	\begin{pmatrix} 0 & \partial_k F({k}) \\ {\partial_k}  F^*({k}) & 0 \end{pmatrix}
	\begin{pmatrix}
	1 \\
	-s' e^{-\im\chi(k)}	
	\end{pmatrix}
\notag\\
	&=\frac{a}{2\varepsilon_+(k)}\left\{\im (s'-s)(g_0^2-g_1^2)-2(s'+s)g_0g_1 \sin(2ka)   \right\}\ (s,\,s'=\pm)~.
	 \end{align*}

For $\sigma^I(\omega)$, we need the matrix element $\langle +|\partial_k\oldhat{\mathcal H}|-\rangle=-\im a (g_0^2-g_1^2)/\varepsilon_+(k)$. After some algebra, by~\eqref{eq:interb-cond-matrix} we obtain (with $b=\pi/a$)
\begin{equation}\label{eq:int-cond-int-BZ}
	\sigma^I(\omega)=\frac{\im 2\sigma_0}{\pi}a^2(g_0^2-g_1^2)^2(\omega+\im\tau^{-1})\inty{0}{b}{\frac{f(\varepsilon_+(k))-f(\varepsilon_-(k))}{4\varepsilon_+(k)^2-(\omega+\im\tau^{-1})^2}\frac{1}{\varepsilon_+(k)^3}}{k}~.
\end{equation}
This representation provides the answer to Problem~1 (Section~\ref{subsec:prob-stat}). The mapping $k\mapsto z$ with $z=e^{2\im ka}$  yields the formulas displayed in~\eqref{eqs:result-preview}.

\subsection{Calculation of $\widetilde I(\lambda,\nu,\vartheta)$: Proof of Proposition~\ref{prop:exact_MT}}
\label{subsec:MellinT}
Next, we compute $\widetilde I(\lambda,\nu,\vartheta)$ by starting with~\eqref{eq:I-MT} in view of definition~\eqref{eq:I-int-def}. We first carry out each of the 1D integrations with respect to $\epsilon_j$ ($j=3,\,1,\,2$) in a specific order, treating $r(z)$ as a nonzero and finite parameter; and finally integrate along the unit circle in the $z$-plane. In the course of this procedure, we determine the region $\mathbb{D}$ via integrability requirements. Without loss of generality, we treat all $\epsilon_j$ as positive ($\epsilon_j\geq 0$, $\forall j$).

Let us focus on the temperature-related integral
\begin{align*}
	\widetilde I_3(\vartheta):=&\inty{0}{\infty}{\epsilon_3^{-\vartheta}\,\{\breve f(z;\epsilon_1, \epsilon_3)-\breve f(z;\epsilon_1, -\epsilon_3)\}}{\epsilon_3}\notag\\
	=&-\{\epsilon_1+r(z)\}^{\vartheta-1}\lim_{w\to -1\atop |w|<1}\inty{0}{\infty}{\epsilon^{-\vartheta}\frac{1-e^{-\epsilon}}{1-we^{-\epsilon}}}{\epsilon}~.
\end{align*}
This integral converges and the requisite limit, as $w\to -1$ within the unit disk, exists if $1< \Re\vartheta< 2$ which contributes to determining region $\mathbb{D}$. By expansion of $(1-we^{-\epsilon})^{-1}$ in powers of $we^{-\epsilon}$ and term-by-term integration, we find
\begin{equation*}
	\widetilde I_3(\vartheta)=\{\epsilon_1+r(z)\}^{\vartheta-1}\Gamma(1-\vartheta)\biggl\{1+\lim_{w\to -1\atop |w|<1}\sum_{n=1}^\infty w^n \left[(n+1)^{\vartheta-1}-n^{\vartheta-1}\right]  \biggr\}~.
\end{equation*}
We now invoke the known function $\Phi(w,s,v):=\sum_{n=0}^\infty (v+n)^{-s} w^n$ as $w\to -1$ with $|w|<1$, while $s=1-\vartheta$, $v=1$; and Joncqui\`ere's relation, viz.,~\cite{Bateman-I}
\begin{equation*}
	L(w,s)+e^{\im s\pi} L(1/w,s)=\frac{(2\pi)^s}{\Gamma(s)} e^{\frac{\im \pi s}{2}} \,\zeta\biggl(1-s, \frac{\ln w}{2\pi \im}\biggr)~,\ L(w,s):=w\Phi(w,s,1)~,
\end{equation*}
where $0< \mathrm{Arg}(w-1)< 2\pi$ and $\zeta(\vartheta,\varsigma)=\sum_{n=0}^\infty (\varsigma+n)^{-\vartheta}$ is the generalized zeta function for $\Re\vartheta >1$, $-\varsigma\notin \mathbb{N}$  (Appendix~\ref{app:zeta}). Thus, we obtain
\begin{subequations}\label{eqs:tilde-integrals}
\begin{equation}\label{eq:tilde-I3}
	\widetilde I_3(\vartheta)=-(2\pi)^{1-\vartheta}\{\epsilon_1+r(z)\}^{\vartheta-1}\frac{\zeta\bigl(\vartheta, \frac12\bigr)}{\sin(\pi\vartheta/2)}~.
\end{equation}

The next task is to compute~\cite{Bateman-I}
\begin{align}\label{eq:tilde-I1}
	\widetilde I_1(\lambda,\vartheta):=&\inty{0}{\infty}{\epsilon_1^{-\lambda}\{\epsilon_1+r(z)\}^{\vartheta-\frac52}}{\epsilon_1}\notag\\
	=& r(z)^{-\lambda+\vartheta-\frac32}\,\frac{\Gamma(1-\lambda)\,\Gamma\bigl(\lambda-\vartheta+\frac32\bigr)}{\Gamma\bigl(\frac52-\vartheta \bigr)}~.
\end{align}
Evidently, the integral $\widetilde I_1(\lambda,\vartheta)$ converges if $\Re(\vartheta-\lambda)<3/2$ and $\Re\lambda <1$.

Regarding the integration with respect to $\epsilon_2$, we have~\cite{Bateman-I}
\begin{equation}
	\widetilde I_2(\nu):=\inty{0}{\infty}{\epsilon_2^{-\nu}\{\epsilon_2+r(z)\}^{-1}}{\epsilon_2}=r(z)^{-\nu}\Gamma(1-\nu)\,\Gamma(\nu)~.
\end{equation}
\end{subequations}
This integral converges provided $0<\Re\nu<1$.

\begin{figure}
  \centering
\includegraphics[scale=.35, trim=0in 0.5in 0in 0in]{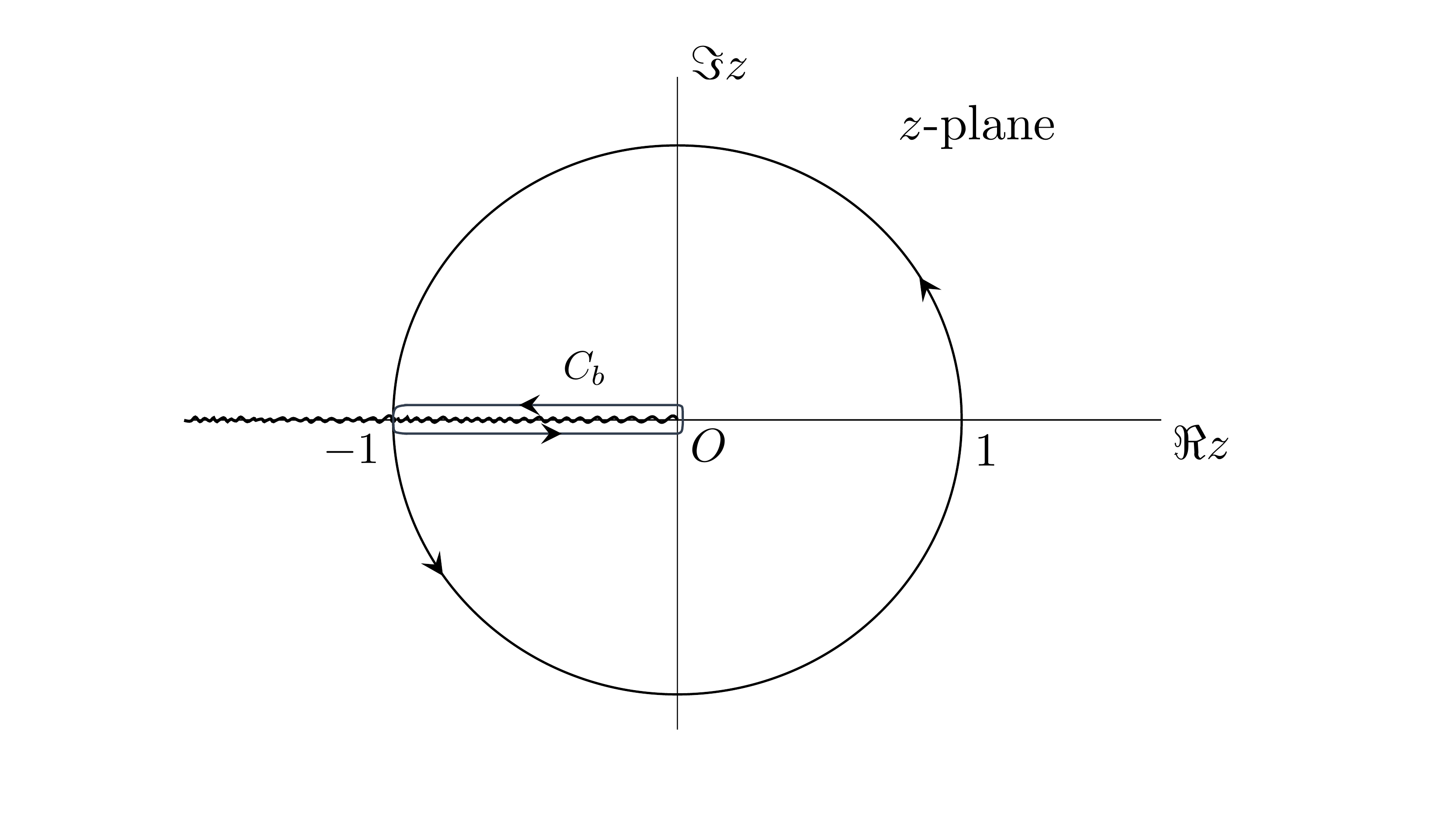}
  \caption{%
    Contours for integral $\Xi(\varsigma)$ and requisite branch cuts (wavy curves) in the $z$-plane. The unit circle (initial integration path) is deformed to the contour $C_b$ which is wrapped around the branch cut from $-1$ to $0$.}
  \label{fig:contour}
\end{figure}

By combining the above results, we write
\begin{equation*}
	\widetilde I(\lambda,\nu,\vartheta)=-(2\pi)^{1-\vartheta}\frac{\zeta\bigl(\vartheta,\frac12\bigr)}{\sin(\pi\vartheta/2)}\frac{\Gamma(1-\lambda)\Gamma\bigl(\lambda-\vartheta+\frac32\bigr)\Gamma(1-\nu)\Gamma(\nu)}{\Gamma\bigl(\frac52-\vartheta\bigr)}\,\Xi(\lambda+\nu-\vartheta)
\end{equation*}
where
\begin{equation*}
	\Xi(\varsigma):=\frac{1}{2\pi\im} \oint_{\{|z|=1\}} z^{\varsigma+\frac12} (1+z)^{-2\varsigma-3}\,\db z~.
\end{equation*}
Our task now is to compute $\Xi(\varsigma)$. This integral converges if $\Re(2\varsigma+3)<1$ which implies $\Re(\lambda+\nu)< \Re\vartheta-1$. Evidently, the integrand has branch points at $z=0$ and $z=-1$. It can be shown that the associated cuts can be defined as separate line segments from $-\infty$ to $-1$ and from $-1$ to $0$ in the real axis (see Fig.~\ref{fig:contour}). By deforming the initial integration path (unit circle) to the contour $C_b$, as depicted in Fig.~\ref{fig:contour}, we find the alternate representation
\begin{align}\label{eq:I-fin-int-MT}
\Xi(\varsigma)&=\frac{1}{2\pi\im} \oint_{C_b} z^{\varsigma+\frac12} (1+z)^{-2\varsigma-3}\,\db z\notag\\
&=	\frac{1}{2\pi\im}\int_0^1 x^{\varsigma+\frac12} (1-x)^{-2\varsigma-3}\left\{e^{-\im\pi\left(\varsigma+\frac12\right)}- e^{\im\pi\left(\varsigma+\frac12\right)}\right\}\notag\\
   &=-\frac{1}{\pi}\cos(\varsigma\pi)\,\frac{\Gamma\bigl(\varsigma+\frac32\bigr)\,\Gamma(-2\varsigma-2)}{\Gamma\bigl(-\varsigma-\frac12\bigr)}~.
\end{align}

Let us collect all the integration results pertaining to $\widetilde I(\lambda,\nu,\vartheta)$. After some algebra by use of~\eqref{eqs:tilde-integrals} and~\eqref{eq:I-fin-int-MT} along with the known identities~\cite{Bateman-I}
\begin{equation*}
\Gamma(-2\varsigma-2)=\frac{1}{\sqrt{\pi}}2^{-2\varsigma-3} \Gamma(-\varsigma-1)\,\Gamma\bigl({\textstyle -\varsigma-\frac12}\bigr)~,\ \cos(\varsigma\pi)=\frac{\pi}{\Gamma\bigl(\frac12+\varsigma\bigr)\,\Gamma\bigl(\frac12-\varsigma\bigr)}~,	
\end{equation*}
we obtain~\eqref{eq:I-MT-exact}. The description of region $\mathbb{D}$ follows from the above regions of integral convergence. This concludes the proof of Proposition~\ref{prop:exact_MT}. \hfill $\square$

\subsection{Zero-temperature limit of $\sigma^I$ and proof of Proposition~\ref{prop:zeroT-I}}
\label{subsec:zero-T}
Next, we define the zero-temperature limit of $\sigma^I(\omega)$ via $\widetilde I(\lambda, \nu,\vartheta)$; see~\eqref{eq:I-MT-exact}. Then we use this definition along with inversion formula~\eqref{eq:I-MT-inv} to derive approximation~\eqref{eq:I-asympt-zeroT} of Proposition~\ref{prop:zeroT-I}. The  condition $\epsilon_3\sqrt{\epsilon_1}\gg 1$, assumed in Proposition~\ref{prop:zeroT-I}, underlies our procedure but is not explicitly invoked in our proof. This condition is discussed in Section~\ref{subsec:nonzero-T-rem}. \color{black}

First, let us recall that taking the limit as $T\to 0$ of the optical conductivity, while keeping all other parameters fixed, formally means setting $f(\varepsilon_+(k))\equiv 0$ and $f(\varepsilon_-(k))\equiv 1$ in momentum integral~\eqref{eq:int-cond-int-BZ}~\cite{AshcroftMermin-book}; cf. Section~\ref{subsec:prob-stat}. The resulting integral for $\sigma^I$ is convergent. We need to define the inversion of $\widetilde I(\lambda, \nu, \vartheta)$ in terms of iterated integrals with respect to the dual variables in a fashion \emph{consistent} with the above formal limit as $\epsilon_3\to \infty$.

For this purpose, we write~\eqref{eq:I-MT-inv} as
\begin{subequations}\label{eqs:INV-MT-mod}
\begin{equation}\label{eq:I-MT-inv-mod}
I(\epsilon_1,\epsilon_2,\epsilon_3)=\frac{1}{(2\pi\im)^2}\int\limits_{\mathbb{C}_2} \db (\lambda,\nu)\,\epsilon_2^{\nu-1}\epsilon_1^{\lambda-1}\left\{\frac{1}{2\pi \im}
\int\limits_{\gamma_3-\im\infty}^{\gamma_3+\im \infty}\widetilde I(\lambda,\nu,\vartheta)\,\epsilon_3^{\vartheta-1}\,\db\vartheta\right\}~. 	
\end{equation}
Here, $\int_{\mathbb{C}_2} \db(\lambda, \nu)$ denotes an appropriate  integral with respect to the complex variables $\lambda$ and $\nu$ over some region $\mathbb{C}_2\subset \mathbb{C}^2$. This $\mathbb{C}_2$ and the real constant $\gamma_3$ ($\gamma_3\in\mathbb{R}$) are subject to restrictions according to the definition of region $\mathbb{D}$ (Proposition~\ref{prop:exact_MT}). For our purposes, we define
\begin{equation}\label{eq:C2-def}
	\mathbb{C}_2:=\{(\lambda, \nu) : \Re\lambda=\gamma_1,\, \Re\nu=\gamma_2;\, -\textstyle{\frac12}<\gamma_1 < 0,\, 0<\gamma_2<\textstyle{\frac12}, \gamma_1+\gamma_2<0\}~;
\end{equation}
\end{subequations}
hence, $1<\gamma_3< \min(3/2+ \Re\lambda, 2)$ in view of $\mathbb{D}$. It is of essence to allow $\gamma_3=\Re\vartheta$ to have greatest lower bound equal to $1$ in the integrand of~\eqref{eq:I-MT-inv-mod}.

\begin{definition}[Zero-temperature limit of $\sigma^I(\omega)$]
	Consider inversion formula~\eqref{eq:I-MT-inv-mod}, subject to~\eqref{eq:C2-def} and $1<\gamma_3< \min(3/2+ \Re\lambda, 2)$. By~\eqref{eq:sigmaI-prev}, the limit of $\sigma^I(\omega)$ as $T\to 0$ comes from the residue of $\widetilde I(\lambda,\nu,\vartheta)\,\epsilon_3^{\vartheta-1}$ at the simple pole $\vartheta=\vartheta_p=1$ in the iterated integral with respect to $\vartheta$. Recall that $\vartheta_p$ coincides with the pole of the Riemann zeta function, $\zeta(\vartheta)$, in $\widetilde I(\lambda,\nu,\vartheta)$. \label{def:zero-T}
\end{definition}

We can verify that Definition~\ref{def:zero-T} yields the expected integral formula for $\sigma^I(\omega)$ over the Brillouin zone, with $f(\varepsilon_+(k))\equiv 0$ and $f(\varepsilon_-(k))\equiv 1$.  By inspection of~\eqref{eq:I-MT-exact}, for fixed $\lambda$ and $\nu$, the simple pole of $\widetilde I(\lambda,\nu, \vartheta)$ at $\vartheta=1$ is the pole of the generalized zeta function $\zeta(\vartheta, 1/2)=(2^\vartheta-1)\zeta(\vartheta)$; cf.~\eqref{eq:tilde-I3} and Appendix~\ref{app:zeta}. Next, we use Definition~\ref{def:zero-T} in order to prove Proposition~\ref{prop:zeroT-I}.

\subsubsection{Proof of Proposition~\ref{prop:zeroT-I}}
\label{sssec:proof-Prop2}
By calculation of the residue at $\vartheta=1$ of $\widetilde I(\lambda,\nu,\vartheta)\,\epsilon_3^{\vartheta-1}$, while $\lambda$ and $\nu$ are held fixed, we have
\begin{equation*}
 I\big|_{T=0}= -\frac{1}{\pi}\frac{1}{(2\pi\im)^2}\int\limits_{\mathbb{C}_2}\db(\lambda,\nu)\,\epsilon_1^{\lambda-1}\epsilon_2^{\nu-1}\frac{\Gamma(1-\lambda)\Gamma\bigl(\lambda+{\textstyle\frac12}\bigr)\Gamma(1-\nu) \Gamma(\nu)\Gamma(-\nu-\lambda)}{2^{2(\nu+\lambda)}\Gamma\bigl({\textstyle \frac12}-\lambda-\nu\bigr)}.
\end{equation*}
Recall that $\epsilon_1\ll 1$. We integrate in $\lambda$ while keeping $\nu$ fixed and enforcing $-1/2<\Re\lambda < -\Re\nu$; cf.~\eqref{eq:C2-def}. Hence, we shift the integration path to the right of the above strip in the $\lambda$-plane, and pick up only the residue of $\Gamma(-\lambda-\nu)$ at $\lambda=-\nu$. For the moment, let us neglect contributions from poles in the region $\{\Re\lambda > -\Re\nu\}$, since these yield higher powers of $\epsilon_1$. We will see that the ensuing approximation for $I$ captures the singularity of $\sigma^I(\omega)$ at the resonance of the smallest bandgap; but needs to be improved.

Considering only the pole at $\lambda=-\nu$ (with fixed $\nu$), we compute
\begin{equation*}
	I\big|_{T=0}\sim -\frac{\pi^{-3/2}}{\epsilon_1\epsilon_2}\frac{1}{2\pi \im}\int\limits_{\gamma_2-\im \infty}^{\gamma_2+\im \infty}\db \nu \,\left(\frac{\epsilon_2}{\epsilon_1} \right)^\nu\,\Gamma(1-\nu)\, \Gamma(\nu)\, \Gamma(1+\nu)\, \Gamma\bigl({\textstyle\frac12}-\nu \bigr)=:I^{(0)}
\end{equation*}
for $\epsilon_1\ll 1$, where $0<\gamma_2< 1/2$. We carry out the contour integral for $I^{(0)}$ exactly in terms of the hypergeometric function ${}_2F_1$~\cite{Bateman-I}, by shifting the integration path to the left of the strip
$\{0<\Re\nu< 1/2\}$ in the $\nu$-plane, for $0<|\epsilon_1/\epsilon_2|< 1$. Thus, we pick up the residues of the integrand at the simple pole located at $\nu=0$ and the double poles at $\nu=-n$, $n\in \mathbb{N}\setminus\{0\}$. After some algebra, this procedure leads to the convergent series
\begin{equation*}
	I^{(0)}= -\frac{1}{\pi}\frac{1}{\epsilon_2^2}\left\{ \frac{\epsilon_2}{\epsilon_1}-\frac{1}{\sqrt{\pi}}\sum_{n=0}^\infty \left(\frac{\epsilon_1}{\epsilon_2}\right)^n \frac{\Gamma\bigl({\textstyle\frac32}+n \bigr)}{n!} \left[\psi(1+n)-\psi\bigl({\textstyle\frac32}+n \bigr)-\ln\biggl(\frac{\epsilon_1}{\epsilon_2}\biggr)\right]\right\}
\end{equation*}
where $\psi(z):=\frac{\db}{\db z}\Gamma(z)$ here. By manipulating this series, we find
\begin{subequations}\label{eqs:I-leading}
\begin{align}\label{eq:I-leading}
I^{(0)}(\epsilon_1, \epsilon_2, \epsilon_3)&= -\frac{1}{\pi}\frac{1}{\epsilon_1 \epsilon_2}\left\{1-\frac{\epsilon_1}{3\epsilon_2}\,{}_2F_1\bigl({\textstyle\frac32},1; {\textstyle\frac52}; 1-\epsilon_1/\epsilon_2\bigr)  \right\}~.	
\end{align}
The requisite hypergeometric function can be computed as
\begin{equation}\label{eq:2F1-log}
	{}_2F_1\bigl({\textstyle\frac32},1; {\textstyle\frac52}; z\bigr)= \frac{3}{z}\left\{\frac{1}{\sqrt{z}} \,\ln\biggl(\frac{1+\sqrt{z}}{\sqrt{1-z}} \biggr) -1\right\}~.
\end{equation}
\end{subequations}
Details of this calculation are provided in Appendix~\ref{app:hyperg}.
The above result for $I^{(0)}$ is analytically continued to all physically relevant complex $\epsilon_2/\epsilon_1$.

Equation~\eqref{eqs:I-leading} describes the singularity of $\sigma^I(\omega)$ in correspondence to the small bandgap, when $\omega+\im \tau^{-1}\to  2(g_0-g_1)$ (for complex $\omega$) or $\epsilon_2\to 0$. This asymptotic formula for $I\big|_{T=0}$ does not capture the \emph{in principle weaker but physically distinct} resonance of the largest bandgap, as $\epsilon_2 \to -4$, or, $\omega+\im \tau^{-1}\to 2(g_0+g_1)$. A remedy is to include the contributions of the poles at $\lambda= -\nu+1,\,1$ in the approximate calculation of the $\lambda$-iterated integral.

 In this vein, let $I^{(j)}$ denote the contribution to $I\big|_{T=0}$ from the pole at $\lambda=-\nu+1$ (if $j=1$) or $\lambda=1$ ($j=2$). In a way similar to the calculation for $I^{(0)}$, for $0<\gamma_2 <1/2$ we obtain
 \begin{align}\label{eq:I1}
 I^{(1)}&=-\frac{1}{8\pi^{3/2}}\ \frac{1}{\epsilon_2}\ \frac{1}{2\pi\im}\int\limits_{\gamma_2-\im\infty}^{\gamma_2+\im\infty}\db\nu\, \left(\frac{\epsilon_2}{\epsilon_1} \right)^\nu\,\Gamma(1-\nu)\,\Gamma(\nu)^2\,\Gamma\bigl({\textstyle\frac32}-\nu \bigr)	\notag \\
 &=-\frac{1}{8\pi\sqrt{\pi}} \frac{1}{\epsilon_2}\sum_{n=0}^\infty \left(\frac{\epsilon_1}{\epsilon_2} \right)^\nu \frac{\Gamma\bigl(n+{\textstyle\frac{3}{2}} \bigr)}{n!}\left\{\psi(1+n)-\psi\bigl({\textstyle\frac32}+n \bigr)-\ln\biggl(\frac{\epsilon_1}{\epsilon_2}\biggr) \right\} \notag\\
 &= -\frac{1}{24\pi}\ \frac{1}{\epsilon_2} \ {}_2F_1\bigl({\textstyle \frac32}, 1; {\textstyle \frac52}; 1-\epsilon_1/\epsilon_2 \bigr)~,
 \end{align}
 \begin{subequations}\label{eqs:I2s}
 \begin{align}\label{eq:I2}
 I^{(2)}&=-\frac{1}{8\sqrt{\pi}}\ \frac{1}{\epsilon_2}\ \frac{1}{2\pi\im}\int\limits_{\gamma_2-\im\infty}^{\gamma_2+\im\infty}\db\nu\, \left(\frac{\epsilon_2}{4} \right)^\nu\,\frac{\Gamma(1-\nu)\,\Gamma(\nu)\,\Gamma(-\nu-1)}{\Gamma\bigl(-{\textstyle\frac12}-\nu \bigr)}	\notag \\
 &=\frac{1}{16\pi}\frac{1}{\epsilon_2} \left\{ \left(\frac{8}{\epsilon_2}-1\right)\ln\biggl(\frac{16}{\epsilon_2}\biggr)+1-\biggl(\frac{8}{\epsilon_2}\biggr)^2 \Gamma\big({\textstyle \frac{3}{2}}\big)\sum_{n=0}^\infty \left(-\frac{4}{\epsilon_2} \right)^n  \frac{\Gamma(1+n)^2}{\Gamma\bigl(n+{\textstyle\frac{3}{2}} \bigr)}\frac{1}{n!}\right\}\notag\\
 &= \frac{1}{16\pi}\frac{1}{\epsilon_2} \left\{ \left(\frac{8}{\epsilon_2}-1\right)\ln\biggl(\frac{16}{\epsilon_2}\biggr)+1-\biggl(\frac{8}{\epsilon_2}\biggr)^2 {}_2F_1\bigl(1, 1;{\textstyle \frac32}; -4/\epsilon_2 \bigr)\right\}~.
 \end{align}
 A few comments on these steps are in order. Regarding $I^{(1)}$, we calculated the residues of the integrand at the double poles  located at $\nu=-n$ ($n\in \mathbb{N}$), thus using the same series as the one involved in $I^{(0)}$. For $I^{(2)}$, we evaluated the residues at the double poles at $\nu=0,\,-1$ and the simple poles at $\nu=-n$, $n\in \mathbb{N}\setminus\{0, 1\}$. We now employ the formula (see Appendix~\ref{app:hyperg})
 \begin{equation}\label{eq:I2-F}
 	{}_2F_1\bigl(1, 1;{\textstyle \frac32}; z\bigr)=\frac{1}{\sqrt{1-z}}\,\frac{\sin^{-1}\bigl(\sqrt{z}\bigr)}{\sqrt{z}}~.
 \end{equation}
\end{subequations}
This function exhibits a singularity that corresponds to the large-bandgap resonance of $\sigma^I(\omega)$, in the limit $\epsilon_2\to -4$ (or $z\to 1$).

Finally, we need to write
	$I\big|_{T=0}\sim  I^{(0)}+I^{(1)}+I^{(2)}$, combining~\eqref{eqs:I-leading}--\eqref{eqs:I2s}. The resulting, modified formula for $I$ yields~\eqref{eq:I-asympt-zeroT} after the  neglect of subdominant terms given that $\epsilon_1\ll 1$, and $|\epsilon_2|\ll 1$ near the first resonance. \color{black} This consideration concludes the proof of Proposition~\ref{prop:zeroT-I}. \hfill $\square$

\subsubsection{On the 3D Mellin transform and zero-temperature expansion}
\label{sssec:rmks-zeroT}
Our use of the exact 3D Mellin transform of $I(\epsilon_1, \epsilon_2, \epsilon_3)$ points to two issues. First, we should justify our choice of applying the 3D Mellin transform to $I$ instead of the (simpler) alternative of applying the 2D Mellin transform to the zero-temperature limit of $I$. Second, it is useful to discuss an estimate for correction terms to~\eqref{eq:I-asympt-zeroT} that come from residues at other poles in the $\lambda$ dual variable (when $\nu$ is fixed).

Regarding the first issue, the reason for our choice is primarily motivated on mathematical grounds: By using the exact formula for $\widetilde I(\lambda, \nu, \vartheta)$, we were able to show that the expected zero-temperature limit of the conductivity corresponds to the pole of the zeta Riemann function included in $\widetilde I(\lambda, \nu, \vartheta)$. Thus, in this sense, we formally demonstrate the mapping of a \emph{physical} limit
(zero temperature) to a singularity of the respective 3D Mellin transform in the context of the linear optical response theory.

In passing, we are now tempted to ask the following question: Can one utilize the exact formula for $\widetilde I(\lambda, \nu, \vartheta)$ (Proposition~\ref{prop:exact_MT}) to extend the result of Proposition~\ref{prop:zeroT-I} to nonzero small $T$? If $\epsilon_3\gg 1$, we have been unable to obtain a plausible low-temperature expansion of $\sigma^I$ by inversion of $\widetilde I$ via the next-order residue in the $\vartheta$ dual variable (with fixed $\lambda$ and $\nu$), for $\Re\vartheta <1$. The actual expansion should involve the small parameter $e^{-\epsilon_3\sqrt{\epsilon_1}}$, for large enough $\epsilon_3\sqrt{\epsilon_1}$ (Section~\ref{subsec:nonzero-T-rem}). In fact, the 3D Mellin transform is limited by our definition of the parameter $\epsilon_3$. However, it is worth studying whether, in the inversion procedure for $\widetilde I$, one may be able to exactly sum up the power series in $\epsilon_3$ arising from residues in $\{\Re\vartheta> 1\}$, when $\epsilon_3$ is small; and analytically continue the result to $\epsilon_3\gg 1$. This task is not addressed here.

In regard to the second issue, i.e., the effect of ($\epsilon_2$-dependent) \emph{higher-order terms due to the small bandgap} on the zero-temperature expansion for $\sigma^I$, the last stage of our proof in Section~\ref{sssec:proof-Prop2} provides some clues. By the inversion procedure, we realize that away from the resonances such neglected terms cause an $O(\epsilon_1)$ error. Near each resonance, the neglect of such higher-order terms amounts to an error of the order of the small parameter of the corresponding resonance, e.g., an $O(\epsilon_2)$ error for the small-bandgap resonance. In Section~\ref{sec:comp}, we test asymptotic formula~\eqref{eq:I-asympt-zeroT} against the numerical evaluation of integral~\eqref{eq:int-cond-int-BZ} for a wide range of $\omega$ (i.e., $\epsilon_2$).

\subsection{On the role of the small parameter $e^{-\epsilon_3\sqrt{\epsilon_1}}$}
\label{subsec:nonzero-T-rem}
Next, we discuss the effect of small nonzero temperatures. We develop a formal argument for the condition $\epsilon_3\sqrt{\epsilon_1}\gg 1$ (Proposition~\ref{prop:zeroT-I}) if $\omega$ is real and $\tau$ is finite and nonzero. Consider the  integral~\eqref{eq:int-cond-int-BZ} and replace $k$ by $k/a$. By
\begin{equation*}
	f(\epsilon)-f(-\epsilon)=-1+\frac{2 e^{-\beta \epsilon}}{1+e^{-\beta \epsilon}}\qquad (\beta=1/T >0~,\ \epsilon=\epsilon(k)=\epsilon_+(k/a))~,
\end{equation*}
the (properly scaled) correction to the zero-temperature limit of $\sigma^I(\omega)$ is
\begin{align*}
	\check{\sigma}^I(\omega)&:=\frac{\sigma^I(\omega)-(\sigma^I(\omega))\big|_{T=0}}{2\sigma_0 a} \notag \\
	&=\frac{4\im}{\pi}(g_0^2-g_1^2)^2(\omega+\im\tau^{-1}) \int_0^{\frac{\pi}{2}}\frac{e^{-\beta\epsilon(k)}}{1+e^{-\beta \epsilon(k)}}\frac{1}{4\epsilon(k)^2-(\omega+\im\tau^{-1})^2}\frac{\db k}{\epsilon(k)^3}~.
	\end{align*}

 For fixed material parameters and real $\omega$, a baseline estimate for $|\check{\sigma}^I(\omega)|$ can be derived from
\begin{subequations}\label{eqs:R-I-fcns}
\begin{align}\label{eq:R-I-int}
|\check{\sigma}^I(\omega)| \le  \frac{4}{\pi} \frac{(g_0+g_1)^2}{g_0 g_1} (g_0-g_1)^2 \, |\omega+\im \tau^{-1}|\, R_*\ e^{-\beta (g_0-g_1)}  \int_0^{\frac{\pi}{2}} \frac{\db k}{\epsilon(k)^3}~.
\end{align}
In the above, we use  $\epsilon_+(k)\ge g_0-g_1$ for all $k$, and define (by $\epsilon\mapsto x=4\epsilon^2$)
\begin{align}\label{eq:R-I-def}
R_*&:=g_0 g_1\,\max_{\omega\in \mathbb{R}}\max_{x\in\mathbb{I}}\left(R(x; \omega)\right)~,\quad \mathbb{I}:=[4(g_0-g_1)^2, 4(g_0+g_1)^2]~;\notag\\
R(x; \omega)&:= \left\{(x-\omega^2+\tau^{-2})^2+4\omega^2\tau^{-2}\right\}^{-1/2}\quad (x\in \mathbb{I},\ \omega\in \mathbb{R})~.  	
\end{align}
In regard to the integral on the right-hand side of~\eqref{eq:R-I-int}, we have
\begin{equation}
	\int_0^{\frac{\pi}{2}} \frac{\db k}{\epsilon(k)^3}\le \frac{\pi}{2} \frac{c}{\sqrt{g_0 g_1} (g_0-g_1)^2}~,
\end{equation}
\end{subequations}
where $c$ is an immaterial numerical constant ($c>0$) and the factor of $\pi/2$ is included for later algebraic convenience. We obtain this estimate by applying the inequality $\epsilon_+(\pi/2-k)^2\ge (g_0-g_1)^2+(16/\pi^2) g_0 g_1 k^2$ for all $k\in [0, \pi/2]$; and integrating in $k$ by scaling out $g_0-g_1$ via a suitable change of variable.

The next task is to compute the dimensionless quantity $R_*$ by usual calculus methods. This $R_*$ depends on the dimensionless parameters $(g_0-g_1)\tau$ and $\sqrt{g_0 g_1}\tau$. After some manipulations, we obtain the formula
\begin{equation}\label{eq:check-R-def}
R_*= \frac{1}{4}\frac{g_0 g_1\tau }{g_0-g_1}\,\check{R}(2(g_0-g_1)\tau);\quad \check{R}(\xi):= \left\{
\begin{array}{lr}
\displaystyle 	\frac{2\xi}{\xi^2 +1}~, & \mbox{if}\ \  0\le \xi<1~, \cr
\displaystyle 1~, & \mbox{if}\ \ \xi\ge 1~.
\end{array} \right.
\end{equation}
Note that $\check{R}(\xi)$ is bounded and continuously differentiable in $[0, \infty)$.

By combining~\eqref{eqs:R-I-fcns} and~\eqref{eq:check-R-def}, and then writing $\xi=2(g_0-g_1)\tau=2\tau\sqrt{g_0 g_1}\sqrt{\epsilon_1}$ and $|\omega+\im \tau^{-1}|=2\sqrt{g_0 g_1}\sqrt{\epsilon_1-\epsilon_2}$, we derive the estimate
\begin{equation}\label{eq:corr-estimate}
|\check{\sigma}^I(\omega)|\le c \frac{(g_0+g_1)^2\tau}{\sqrt{g_0 g_1}}\ \frac{\sqrt{\epsilon_1-\epsilon_2}}{\sqrt{\epsilon_1}}\check{R}(2\tau\sqrt{g_0 g_1}\sqrt{\epsilon_1})\ e^{-\epsilon_3\sqrt{\epsilon_1}} ~.	
\end{equation}
This inequality can be simplified in the cases with $|\epsilon_2|\ll \epsilon_1\ll 1$ and $\epsilon_1\ll |\epsilon_2|\ll 1$, outlined in Remark~\ref{rmk:sigmaI-behavior}. We leave the details to the reader.

Let us compare~\eqref{eq:corr-estimate} to the zero-temperature formula
\begin{equation*}
\left|\frac{(\sigma^I(\omega))\big|_{T=0}}{2\sigma_0 a}\right|=	\frac{1}{2} \frac{(g_0+g_1)^2}{g_0 g_1}\ \epsilon_1 \sqrt{\epsilon_1-\epsilon_2}\  I(\epsilon_1, \epsilon_2, \infty)~,
\end{equation*}
where $I(\epsilon_1, \epsilon_2, \infty):=\lim_{\epsilon_3\to \infty}I(\epsilon_1, \epsilon_2, \epsilon_3)$ is replaced by asymptotic formula~\eqref{eq:I-asympt-zeroT}. By imposing
\begin{equation*}
 c \frac{(g_0+g_1)^2\tau}{\sqrt{g_0 g_1}}\ \frac{\sqrt{\epsilon_1-\epsilon_2}}{\sqrt{\epsilon_1}}\check{R}(2\tau\sqrt{g_0 g_1}\sqrt{\epsilon_1})\ e^{-\epsilon_3\sqrt{\epsilon_1}} \ll 	\left|\frac{(\sigma^I(\omega))\big|_{T=0}}{2\sigma_0 a}\right|~,
\end{equation*}
according to~\eqref{eq:corr-estimate}, we need to distinguish cases for $\check{R}(\xi)$. This procedure yields the $\omega$-independent condition $e^{-\epsilon_3\sqrt{\epsilon_1}}\ll c_1$ where $c_1$ is a positive numerical constant of the order of unity; thus, $\epsilon_3\sqrt{\epsilon_1}\gg 1$.

In regard to the asymptotic evaluation of the correction term $\breve\sigma^I(\omega)$, one may wonder if it is useful to conveniently employ the Mellin transform with respect to $(\epsilon_1, \epsilon_2, \epsilon_3)$ via the (modified) parameter $\epsilon_3=e^{\beta (g_0-g_1)}$ ($\epsilon_3\gg 1$). We leave this problem unresolved in this paper.

\color{black}

\section{Numerical computations}
\label{sec:comp}
 In this section, we validate our asymptotics for the interband conductivity, $\sigma^I(\omega)$, as a function of frequency $\omega$ via numerical computations. We compare our asymptotic results, particularly the zero-temperature formula \eqref{eq:I-asympt-zeroT} that enters~\eqref{eq:sigmaI-prev}, to numerical evaluations of the requisite  integral~\eqref{eq:int-cond-int-BZ} for small but nonzero temperatures ($\epsilon_3\gg 1$).

 First, we choose convenient units of energy and conductivity. Set $g_0+g_1=1$, which fixes the unit of energy; and take $(2a)\sigma_0=1$ which sets the unit of conductivity. In other words, quantities that have the dimension of energy are scaled by $g_0+g_1$; and $\sigma^I(\omega)$ is naturally scaled by $2\sigma_0 a$.  We numerically determine $\sigma^I(\omega)$ by using~\eqref{eq:sigmaI-prev} and integral~\eqref{eq:int-cond-int-BZ} over the Brillouin zone.

 \begin{figure}[!t]
  \centering
  \subfloat[Real part]{
    \includegraphics[scale=0.62]{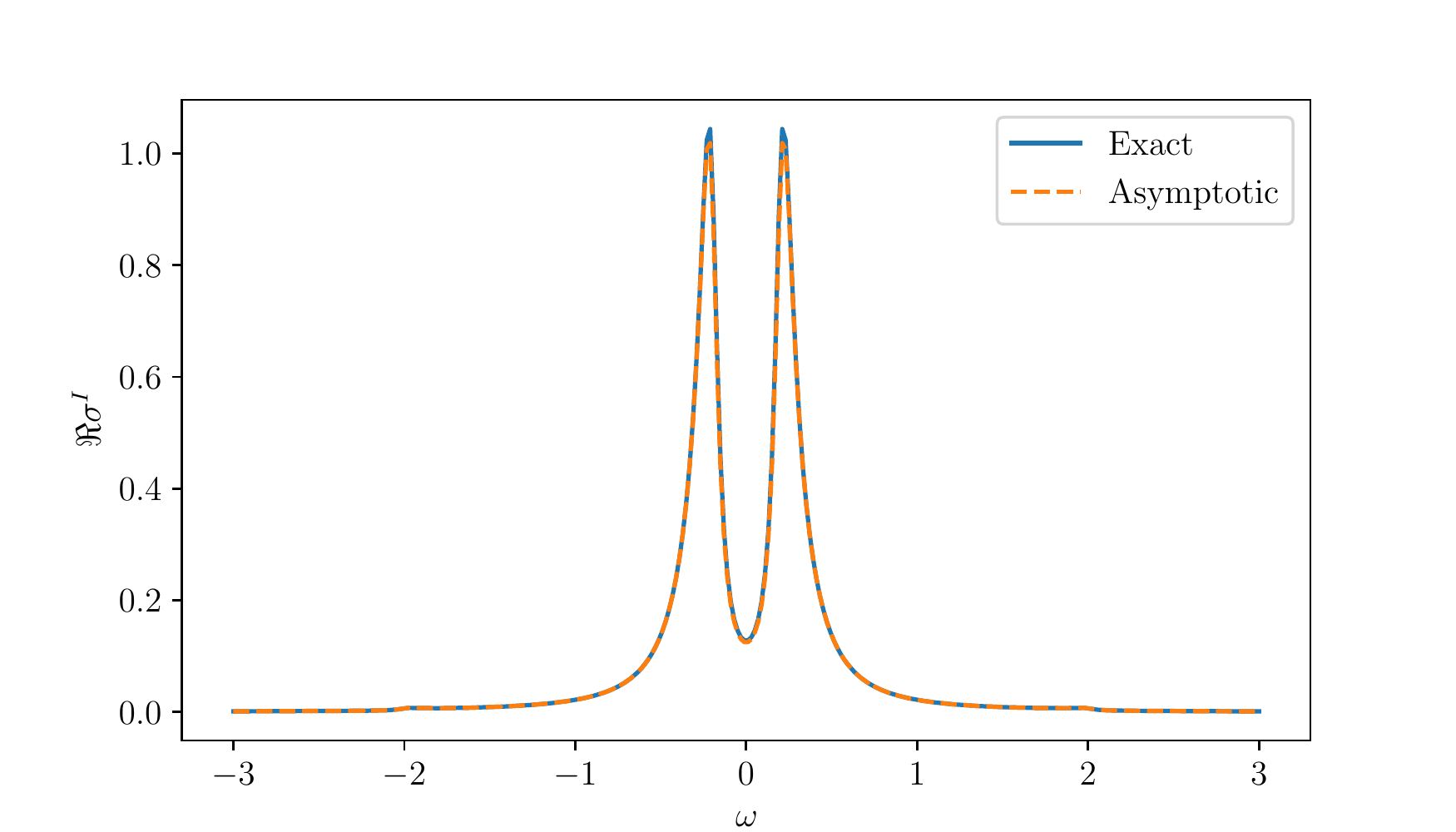}
  }

  \subfloat[Imaginary part]{
    \includegraphics[scale=0.62]{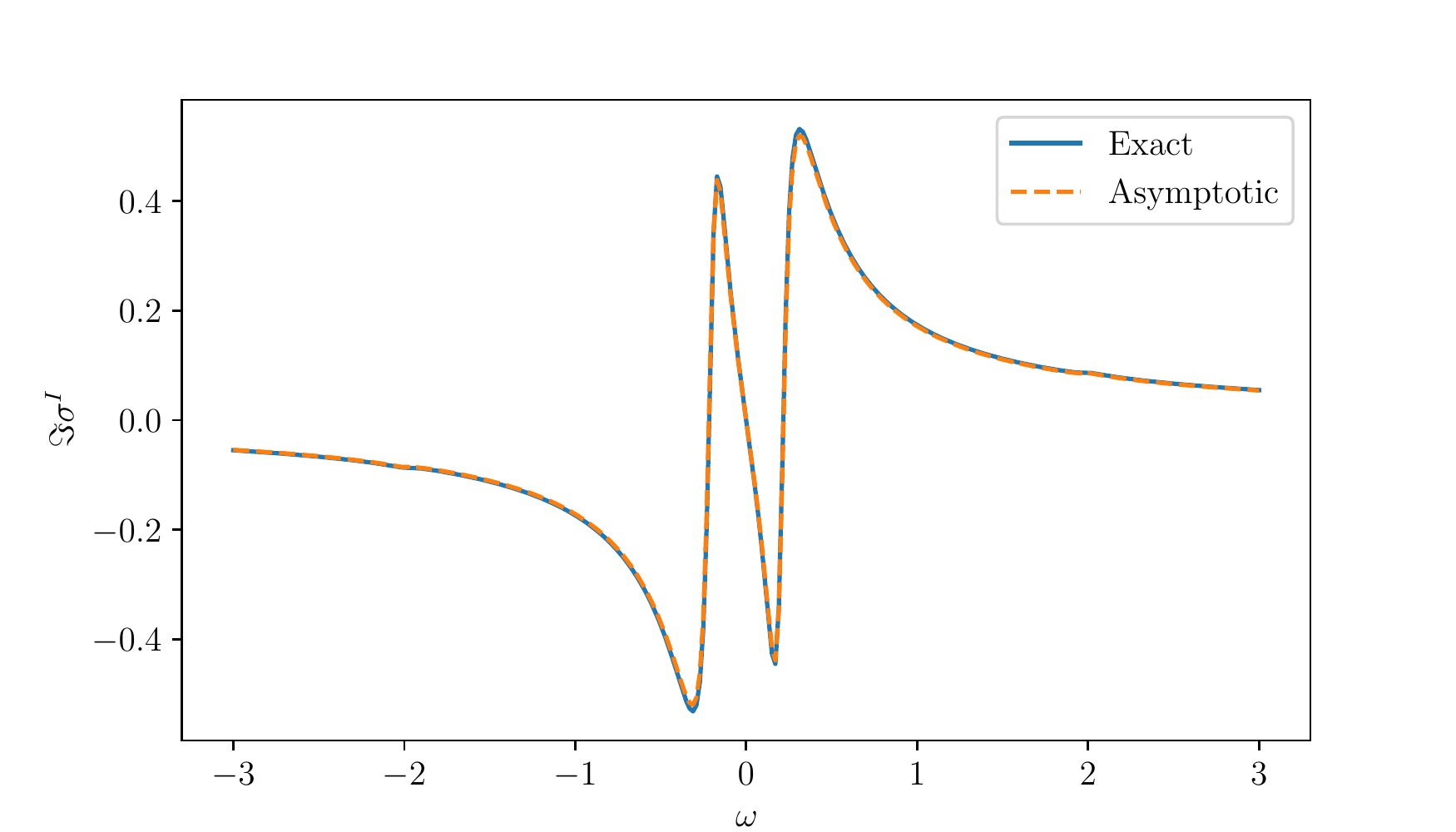}
  }
  \caption{%
  Real part [(a)] and imaginary part [(b)] of $\sigma^I$ versus $\omega$ near zero temperature. The units are such that $g_0+g_1=1$ and $2\sigma_0 a=1$. We use exact formula~\eqref{eq:sigmaI-prev} with~\eqref{eq:int-cond-int-BZ} (solid curve) and asymptotic formula~\eqref{eq:I-asympt-zeroT} (dashed curve). The parameter values are $g_0 = .55$, $\tau^{-1} = 0.05$ and $\beta = 10^{3}$, which give $\epsilon_1 = 0.040$ and $\epsilon_3 = 497$ ($\sqrt{\epsilon_1}\epsilon_3=99.896$); while $\epsilon_2$ varies through $\omega$ with $|\epsilon_2|=0.020$ at the small-bandgap resonance.
    }
  \label{fig:numerical}
\end{figure}

 The real and imaginary parts of $\sigma^I$ are plotted versus $\omega$ in Fig.~\ref{fig:numerical} for $g_0=0.55$, $\tau^{-1}=0.05$ and $\beta=10^3$.  In our numerics, at the smallest-bangap resonance, where $\omega \simeq \pm 2 (g_0 - g_1)$ and $|\epsilon_2(\omega)|$ achieves its minimum with $\omega$, we have $|\epsilon_2| = 0.020$. \color{black} Since we use $\epsilon_1\simeq 0.04$, and $\sqrt{\epsilon_1} \epsilon_3 \simeq 100$, we verify that the plots of Fig.~\ref{fig:numerical} are in the regime of Proposition~\ref{prop:zeroT-I}. We see excellent agreement between the exact integral~\eqref{eq:int-cond-int-BZ} and the asymptotic result for a wide range of $\omega$. Notably, Fig.~\ref{fig:numerical} depicts clearly the small-bandgap resonance.

\begin{figure}
  \centering
  \subfloat[Real part]{
    \includegraphics[scale=.62]{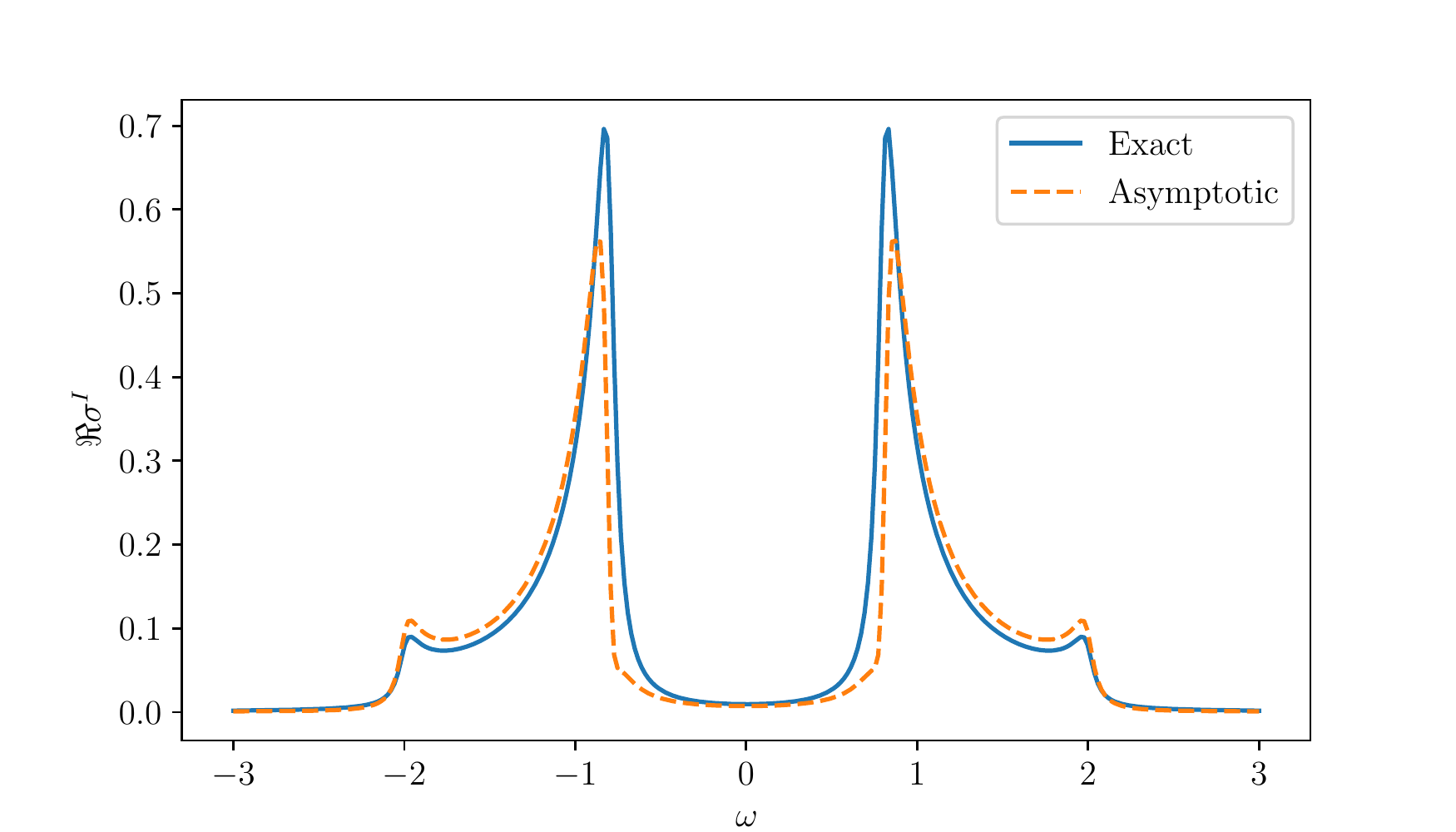}
  }

  \subfloat[Imaginary part]{
    \includegraphics[scale=.62]{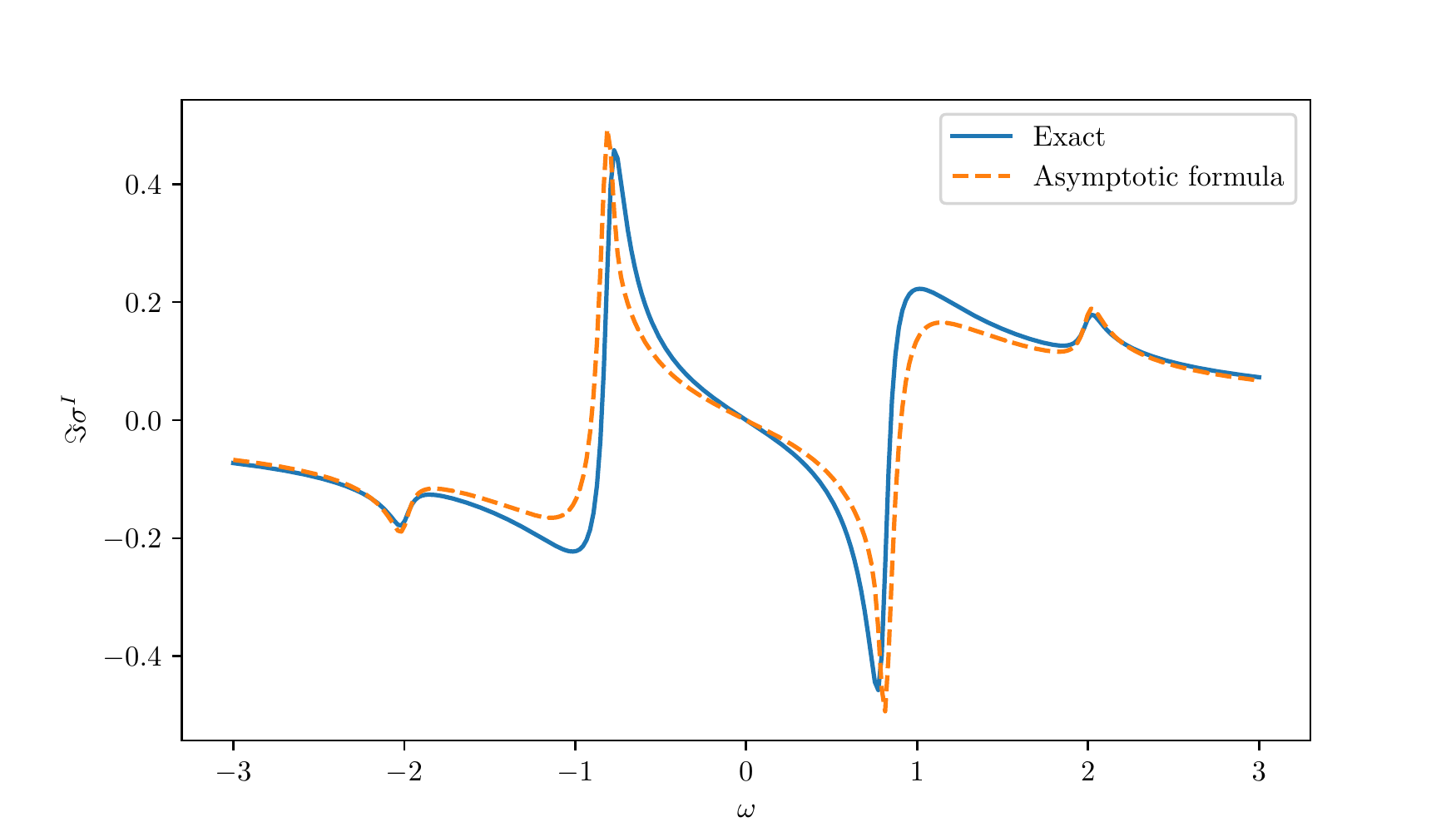}
  }
  \caption{%
      Real part [(a)] and imaginary part [(b)] of $\sigma^I$ versus $\omega$ near zero temperature. The parameter values are $g_0 = 0.7$, $\tau^{-1} = 0.05$ and $\beta = 10^{3}$, which give $\epsilon_1 = 0.762$ and $\epsilon_3 = 458$ ($\sqrt{\epsilon_1} \epsilon_3 = 399.800$); while $\epsilon_2$ varies through $\omega$ with $|\epsilon_2| = 0.095$ at the small-bandgap resonance.
}
  \label{fig:numerical_regimeA_bothresonances}
\end{figure}

Figure~\ref{fig:numerical_regimeA_bothresonances} shows both the small- and large-bandgap resonances clearly, still within the assumed parameter regime. We use the parameters $g_0=0.7$, $\tau^{-1}=0.05$ and $\beta=10^3$. In this case, $\epsilon_1\simeq 0.762$, which slightly spoils the accuracy of asymptotic formula~\eqref{eq:I-asympt-zeroT}, particularly near the highest peak of $\Re\sigma^I(\omega)$. Our asymptotic formula describes both resonances reasonably well.\color{black}

\begin{figure}
  \centering
  \subfloat[Real part]{
    \includegraphics[scale=.62]{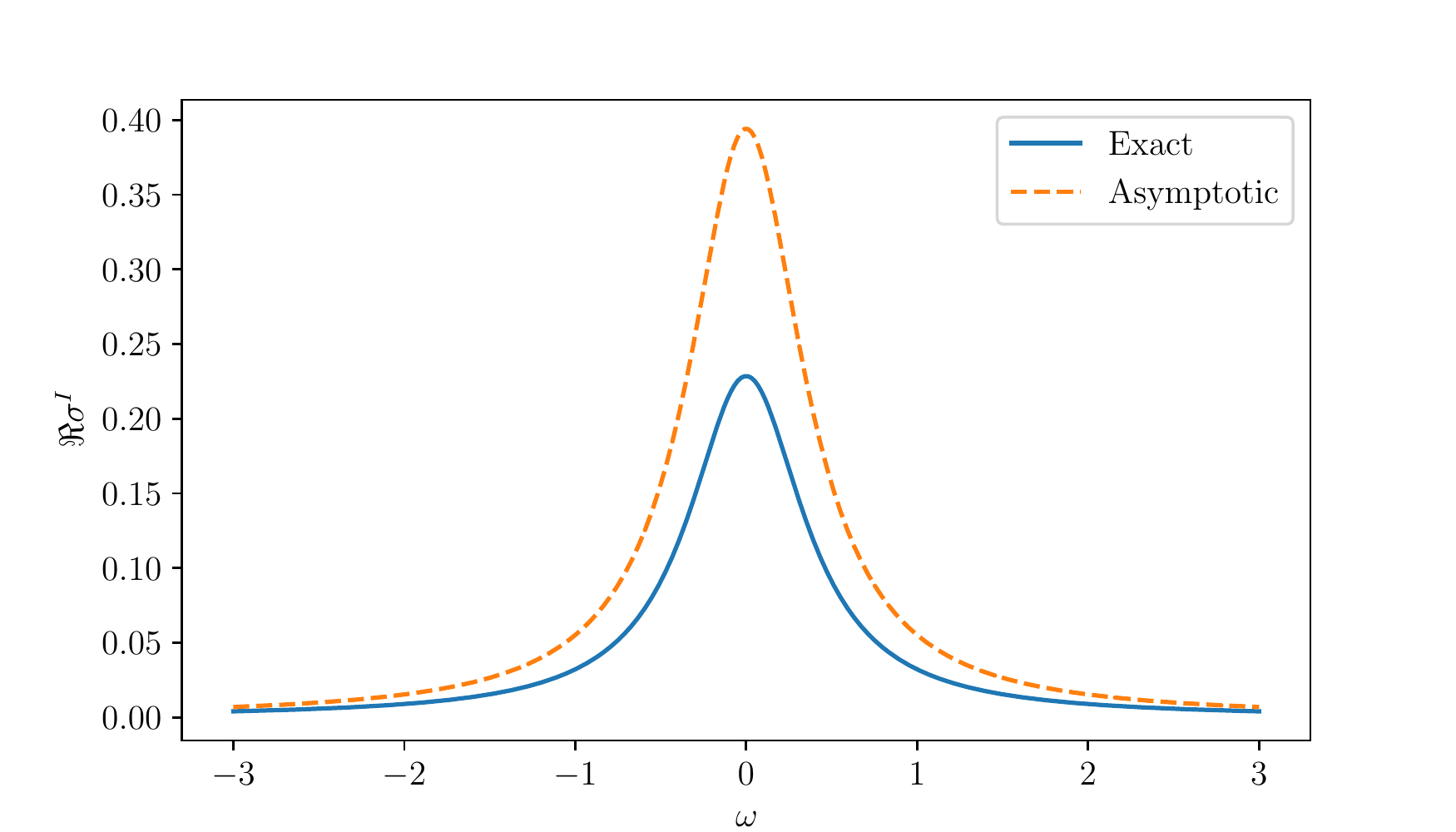}
  }

  \subfloat[Imaginary part]{
    \includegraphics[scale=.62]{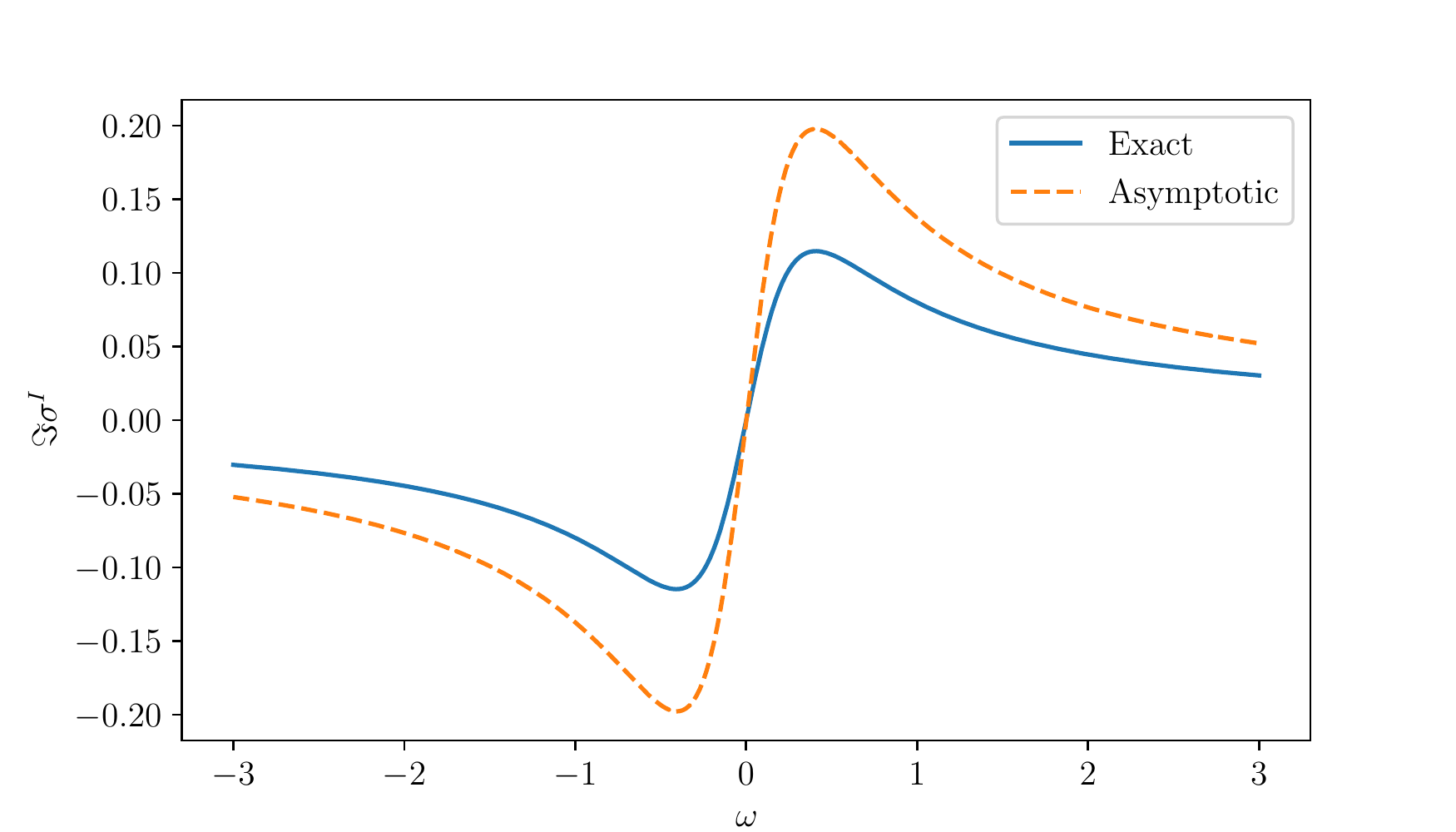}
  }
  \caption{%
      Real part [(a)] and imaginary part [(b)] of $\sigma^I$ versus $\omega$ near zero temperature. The parameter values are $g_0 = .505$, $\tau^{-1} = 0.4$ and $\beta = 100$, which give $\epsilon_1 =4\times 10^{-4}$ and $\epsilon_3 = 49.9$ ($\sqrt{\epsilon_1} \epsilon_3 \simeq 1.00$); while $\epsilon_2$ varies through $\omega$ with $|\epsilon_2| = 0.160$ at the small-bandgap resonance. The second condition (i.e., $\epsilon_3\sqrt{\epsilon_1}\gg 1$) of Proposition~\ref{prop:zeroT-I} is violated. \color{black}
    }
  \label{fig:numerical_regimeA_smallepsilon1epsilon3}
\end{figure}

\begin{figure}
  \centering
  \subfloat[Real part]{
    \includegraphics[scale=.62]{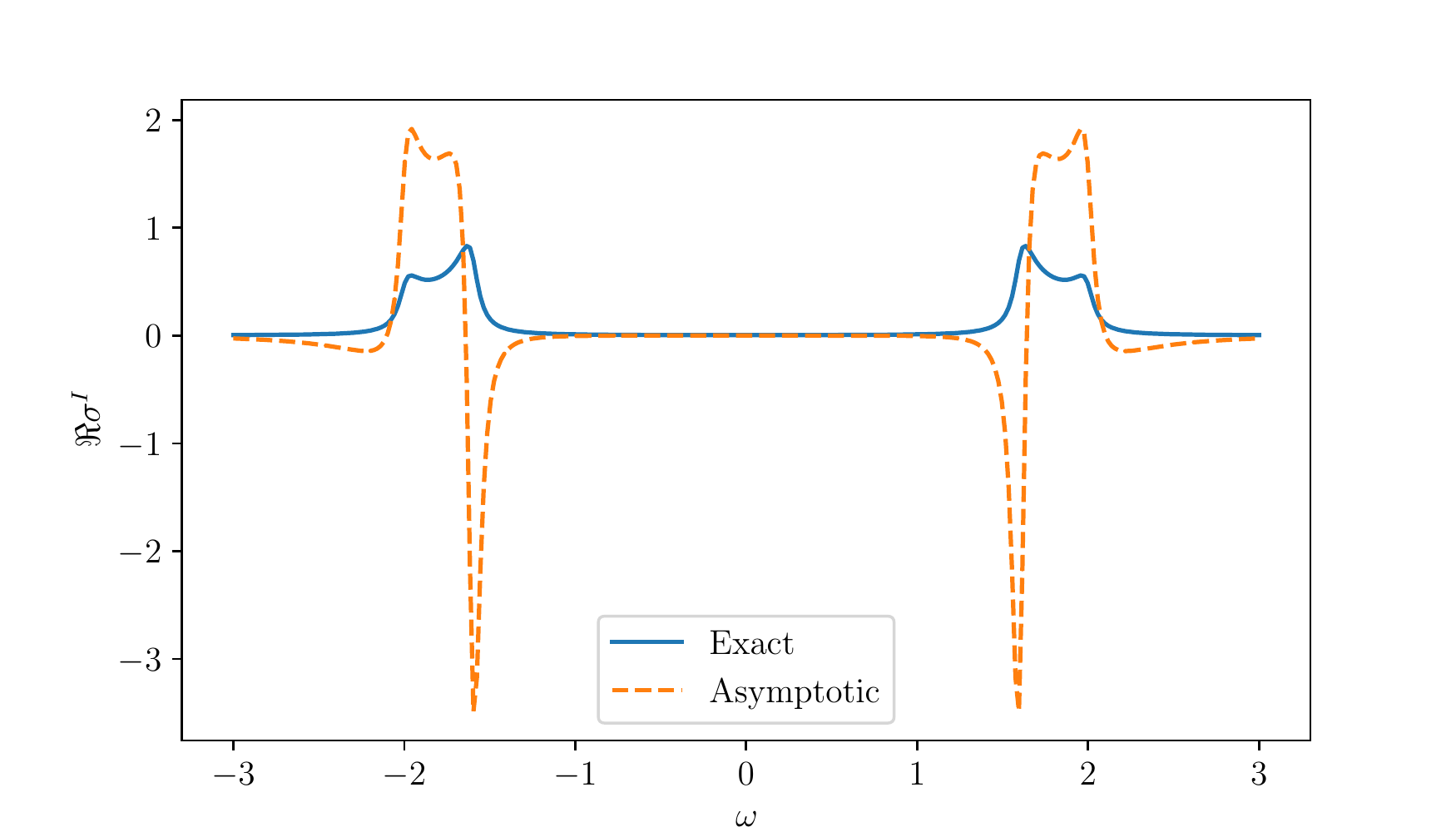}
  }

  \subfloat[Imaginary part]{
    \includegraphics[scale=.62]{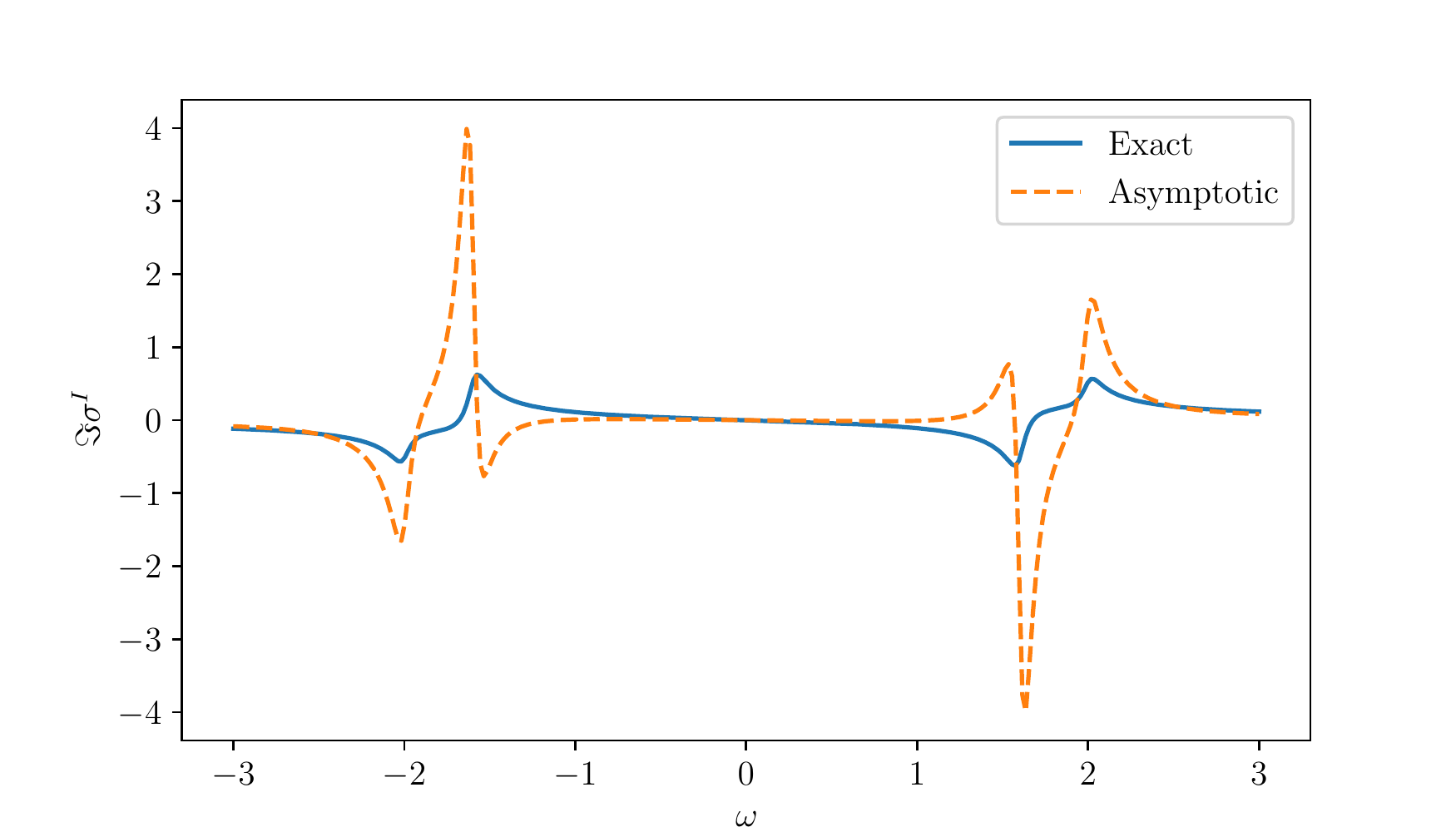}
  }
  \caption{%
     Real part [(a)] and imaginary part [(b)] of $\sigma^I$ versus $\omega$ near zero temperature. The parameter values are $g_0 = 0.900$, $\tau^{-1} = 0.050$ and $\beta = 10^{3}$, which give $\epsilon_1 = 7.11$ and $\epsilon_3 = 300$ ($\sqrt{\epsilon_1} \epsilon_3 = 799.937$); while $\epsilon_2$ varies via $\omega$ with $|\epsilon_2| = 0.444$ at the small-bandgap resonance. The first condition (i.e., $\epsilon_1\ll  1$) of Proposition~\ref{prop:zeroT-I} is violated. \color{black}
      }
  \label{fig:numerical_regimeA_largeepsilon1}
\end{figure}

In contrast, Figs.~\ref{fig:numerical_regimeA_smallepsilon1epsilon3} and~\ref{fig:numerical_regimeA_largeepsilon1} depict cases that are incompatible with the parameter regime of Proposition~\ref{prop:zeroT-I}. Then, our zero-temperature asymptotic formula is inaccurate. For example, the parameter values used in Fig.~\ref{fig:numerical_regimeA_smallepsilon1epsilon3} satisfy  $\epsilon_3\sqrt{\epsilon_1}\simeq 1$, which violates the second condition (i.e., $\epsilon_3\sqrt{\epsilon_1}\gg 1$) of Proposition~\ref{prop:zeroT-I}. Hence, small-temperature effects become important. Figure~\ref{fig:numerical_regimeA_largeepsilon1} provides an example with $\epsilon_1>1$; then, the first condition (i.e., $\epsilon_1 \ll 1$) of Proposition~\ref{prop:zeroT-I} does not  hold. Thus, small-bandgap corrections are significant. \color{black}

\section{Conclusion}
\label{sec:conclusion}
In this paper, we showed that the 3D Mellin transform can be used for the derivation of a zero-temperature asymptotic formula for the interband conductivity, $\sigma^I$, of the 1D SSH model as a function of frequency, $\omega$. This part of the conductivity has an intricate dependence on $\omega$, and in fact exhibits physically appealing resonances at nonzero frequencies. A similar approach can be applied to the intraband conductivity of this model, whose dependence on $\omega$ follows a standard law and, hence, was not addressed here.

Our methodology yields a formula for $\sigma^I(\omega)$ that is valid, for all practical purposes, uniformly in $\omega$ when the bandgap is relatively small and the absolute temperature is sufficiently low. Our approximate, analytical results were found to be in good agreement with direct numerical computations based on the momentum integral for $\sigma^I(\omega)$. We believe that our approach and ensuing result contrasts the traditional point of view on the conductivity which yields local (in $\omega$) information about $\sigma^I(\omega)$ when the material parameters are fixed.
In other words, in our work we focused on approximately deriving $\sigma^I(\omega)$ for a wide range of $\omega$, by assuming that some material parameters take extreme values, i.e., the temperature is low and the bandgap is small. In principle, this methodology can be extended to other parameter regimes of the SSH model, such as the limit of large bandgap (as $\epsilon_1 \rightarrow \infty$).

It is natural to ask whether our approach, relying on the multidimensional Mellin transform technique, can be extended to other more realistic models at low temperatures. Of particular interest are models in higher dimensions when a symmetry of their Hamiltonian is broken so that a nonzero energy bandgap exists. We expect that similar calculations can be carried out for essentially generic systems that exhibit a small bandgap limit. A relatively simple example of such a system, in 2D, is the Haldane model~\cite{Haldane1988}. \color{black}

\section*{Acknowledgments}
The first author (DM) is grateful to the School of Mathematics of the University of Minnesota for hosting him as an Ordway Distinguished Visitor in the spring of 2022, when part of this work was completed.
The research of the second and third authors (ABW and ML) was supported in
part by NSF DMREF Award No. 1922165 and Simons Targeted Grant Award No. 896630.

\appendix 

\section{On a kinetic formulation for linear optical response}
\label{app:conductivity-form}
In this appendix, we review the origin of formula~\eqref{eq:cond_trace}~\cite{Bellissard1998,Bellissard2002,Kubo1957}. Emphasis is placed on the modeling of the energy loss due to electron scattering via the ``relaxation time approximation''~\cite{AshcroftMermin-book}, in which dissipative effects are captured through the effective constant parameter $\tau$. For details from the perspective of partial differential equations, see our expository article~\cite{WML-rev}.

We sketch a derivation of~\eqref{eq:cond_trace} in the spirit of~\cite{Bellissard1998,Bellissard2002}.  Let the unperturbed one-electron Hamiltonian be $\mathcal H$, acting on the Hilbert space $\mathfrak H$. This $\mathcal H$ describes electron motion in $\mathbb{R}^d$ without the electromagnetic field. The time-dependent electric field is $E(\theta+\omega t)$ where $E(\cdot)$ is  $2\pi$-periodic and $\theta$ is a parameter in $[0, 2\pi)$.  The total Hamiltonian is $\mathcal H_E=\mathcal H-e E(\theta+\omega t) \cdot\mathcal X$ where $\mathcal X$ is the position operator.

An observable of interest is the current density operator $\mathcal J=-\im e[\mathcal X, \mathcal H]$, which is proportional to the electron velocity operator. We will define, and describe perturbatively with $E$,
a suitable function, denoted as $\oldhat{J}(\omega)$ ($\oldhat{J}(\omega): \mathbb{C}\rightarrow \mathbb{C}^d$), that results from averaging procedures applied successively to $\mathcal J$.

Now let us recall the notion of the ``density matrix'' operator $\varrho: \mathfrak H\rightarrow \mathfrak H$~\cite{vonNeumann-book}: If a quantum system can occupy any one of the linearly independent (normalized) pure states $\{\psi_j\}_{j\in J}$ prepared with probabilities $\{p_j\}_{j\in J}$ (where $\sum_{j\in J}p_j=1$, $p_j>0$),  the related \emph{ensemble} average, $\langle A\rangle$, of  $\mathcal A: \mathfrak H \rightarrow \mathfrak H$ is~\cite{vonNeumann-book}
\begin{equation}\label{eq:ens-ave}
\langle \mathcal A\rangle :=\sum_{j\in J}p_j \langle\psi_j | \mathcal A \psi_j\rangle_{\mathfrak H}=: \Tr\left\{\mathcal A\varrho\right\}~.
\end{equation}
Here, $\langle \cdot | \cdot \rangle_{\mathfrak H}$ is the inner product on the Hilbert space $\mathfrak H$. Equation~\eqref{eq:ens-ave} suggests $\varrho:=\sum_{j\in J} p_j \mathcal P_j$ where $\mathcal P_j$ is a projector such that $\mathcal P_j \psi_j=\psi_j$~\cite{vonNeumann-book}.

Consider particle motion under the perturbed Hamiltonian, $\mathcal H_E$. The operator $\varrho=\varrho(t;\theta)$ obeys the Liouville-von Neumann evolution equation~\cite{vonNeumann-book,Bellissard1998}, in which the term containing $E(\cdot)$ is treated as a perturbation. This evolution equation can be written as $\frac{\db}{\db t}\varrho(t;\theta)=-\mathcal L_{\mathcal H_E}(\varrho(t;\theta))$ where $\mathcal L_{\mathcal H_E}(\varrho):=\im [\mathcal H_E, \varrho]$ defines the Liouville superoperator.
%
%
Note that $\varrho(t)$ is parametrized by the $\theta$ introduced in the periodic electric field.
%
%

The equation for $\varrho(t;\theta)$ is solved under the following assumptions:

\begin{itemize}

\item The initial condition $\varrho(0^+;\theta)=f(\mathcal H)$ is imposed. Hence, at $t=0$ the electron is at the thermal equilibrium corresponding to the unperturbed Hamiltonian, $\mathcal H$.

\item The collisions of the electron with other particles (e.g., impurities and phonons) occur \emph{instantly} at random times $\{t_n\}_{n=1}^\infty$ where $t_{n+1}>t_n\ge 0$ $\forall n\in \mathbb{N}$. (Set $t_0:=0$.)

\item All differences $\tau_n:=t_{n+1}-t_n>0$ ($\forall \,n\in\mathbb{N}$) are treated as \emph{independent and identically distributed random variables} that follow the Poisson distribution law with parameter $\Gamma_p=\tau^{-1}$; hence,
\begin{equation}\label{eq:P-measure}
	\mathrm{Prob}(\tau_n\le T)=\inty{0}{T}{\Gamma_p\, e^{-\Gamma_p \tau'}}{{\tau'}}\qquad (T>0)~.
\end{equation}

\item Immediately after every collision event, the electron reaches its unperturbed equilibrium; $\varrho(t_n^+;\theta)=f(\mathcal H)$, $\forall n\in\mathbb{N}$. Thus, $\varrho(t;\theta)$ evolves by the Liouville-von Neumann equation for times $t$ in $(t_n, t_{n+1})$, $\forall n\in\mathbb{N}$.

\item The system has vanishing equilibrium current, viz., $\Tr\{\mathcal J f(\mathcal H)\}=0$.

\end{itemize}

These assumptions suggest three types of averages.
First, for fixed $\{t_n\}_{n\in \mathbb{N}}$, one employs the ensemble average of $\mathcal J$, viz.,
\begin{equation}
	\langle \mathcal J(t;\theta)\rangle :=\Tr\left\{\mathcal J \varrho(t;\theta) \right\}\qquad t_n < t<t_{n+1}\quad (\forall n\in\mathbb{N})~.
\end{equation}
Second, by a Tauberian theorem~\cite{Wiener-FT-book}, consider the combined average
\begin{align}
	\langle \mathcal J\rangle_\omega:=&\frac{1}{2\pi}\lim_{t\to +\infty}\frac{1}{t}\int_0^t\!\int_{-\pi}^\pi e^{\im (\omega t'+\theta)}\langle \mathcal J(t';\theta)\rangle\ \db \theta \,\db t' =\frac{1}{2\pi}\lim_{\delta\downarrow 0}\left\{\delta \lim_{N\to \infty}J_N^\delta(\omega)\right\}; \notag\\
	&J_N^\delta(\omega)=\sum_{n=0}^{N-1}\int_{t_n}^{t_{n+1}}\!\int_{-\pi}^\pi e^{-\delta t'}\,e^{\im (\omega t'+\theta)}\langle \mathcal J(t';\theta)\rangle\ \db \theta \,\db t'~.
\end{align}
%
Third, one must account for the randomness of the differences $\{\tau_n\}_{n\in\mathbb{N}}$. Let $\tau^N:=(\tau_0,\,\tau_1,\,\ldots,\,\tau_N)$, and define
\begin{equation}
	\oldhat{J}(\omega):= \mathbb{E}_P\bigl[ \langle \mathcal J\rangle_\omega\bigr]:= \frac{1}{2\pi}\lim_{\delta\downarrow 0}\left\{\delta \lim_{N\to \infty} \mathbb{E}_P[J_N^\delta(\omega; \tau^N)]\right\}~;
\end{equation}
$\mathbb{E}_P[F(\tau^N)]$ is the expectation of $F(\tau^N)$ by the Poisson distribution law.

By linear response, the above steps are complemented with the linearization of  $\varrho(t;\theta)$ in the electric field $E$. The $l$-th component of  $\oldhat{J}$ takes the form
\begin{equation*}
	\oldhat{J}_{l}(\omega)=\sigma_{lm}(\omega)\oldhat{E}_m~,\quad \oldhat{E}_m:=\frac{1}{2\pi}\inty{-\pi}{\pi}{E_m(\theta) e^{\im\theta}}{\theta}\qquad (l,\,m=1,\,\ldots\,,d)~,
\end{equation*}
where $E_m(\theta)$ is the $m$-th component of $E(\theta)$ and $\sigma_{lm}=[\bsig]_{lm}$ is given by~\eqref{eq:cond_trace}. Then, the interband conductivity, $\sigma_{lm}^I(\omega)$, can be obtained from~\eqref{eq:interband-cond}.

\setcounter{equation}{0}

\section{Mellin transform: A review}
\label{app:MT}
In this appendix, we formally review the 1D Mellin transform, and its multidimensional version; see, e.g.,~\cite{Carrier-book,ChengWu-book}. This technique is used in Section~\ref{sec:asymptotics}.

Let us recall the `two-sided' Laplace transform, $\mathfrak L[g]$,  of $g:\mathbb{R}\rightarrow \mathbb{C}$, viz.,
\begin{equation}\label{eq:2-side-LT}
\mathfrak L[g](\nu):=\inty{-\infty}{\infty}{g(t)\, e^{-\nu t}}{t}=:\tilde g(\nu)~,\quad  \breve a< \Re\nu< \breve b~.	
\end{equation}
The restriction on $\Re\nu$ results from the integral convergence requirement, for some real $\breve a$, $\breve b$. We avoid prescribing any (sufficient) conditions on $g$. Typically, $\tilde g(\nu)$ is holomorphic in the strip $\{\nu \in \mathbb{C} : \breve a< \Re\nu< \breve b \}$; but $\tilde g(\nu)$ has singularities in $\{\Re \nu <\breve a\}\cup \{\Re \nu >\breve b\}$. The inverse Laplace transform is
\begin{equation}\label{eq:2-side-LT-INV}
\mathfrak L^{-1}[\tilde g](t):=\frac{1}{2\pi\im}\int\limits_{\breve \gamma-\im \infty}^{\breve\gamma+\im \infty}e^{\nu t}\,\tilde g(\nu)\,\db\nu~,\quad \breve a< \breve\gamma< \breve b~.	
\end{equation}
Under mild conditions on $g(t)$, $\mathfrak L^{-1}[\tilde g](t)=g(t)$ almost everywhere~\cite{Wiener-FT-book,PaleyWiener}. These considerations can be extended to functions $g$ whose domain is $\mathbb{C}$. 

The Laplace transform can of course be generalized to higher dimensions. Given $g: \mathbb{R}^n \rightarrow \mathbb{C}$, one defines $\mathfrak L[g](\nu)$ by the counterpart of~\eqref{eq:2-side-LT} where $t=(t_1,\,\ldots,\,t_n)\in \mathbb{R}^n$, $\nu=(\nu_1,\,\ldots\,, \nu_n)\in \mathbb{C}^n$ and $\nu t$ is replaced by $\nu\cdot t$. The requirement of convergence for the integral implies $(\Re\nu_1,\,\ldots,\,\Re\nu_n) \in \mathbb{D}\subset \mathbb{R}^n$ for some nonempty set $\mathbb{D}$. The $n$-dimensional counterpart of~\eqref{eq:2-side-LT-INV} is
\begin{equation}\label{eq:2-side-LT-INV-n}
\mathfrak L^{-1}[\tilde g](t):=\prod_{j=1}^n \left(\frac{1}{2\pi\im}\int\limits_{\breve\gamma_j-\im\infty}^{\breve\gamma_j+\im \infty}\db\nu_j\  e^{\nu_j t_j}\right) \tilde g(\nu)~,\quad (\breve\gamma_1,\,\ldots,\,\breve \gamma_n)\in\mathbb{D}~.
\end{equation}

The 1D Mellin transform can now be introduced via a nonlinear mapping~\cite{Carrier-book}. In~\eqref{eq:2-side-LT}, map $t \mapsto \wp := e^t$. Hence, with $I(\wp):= g(t(\wp))/\wp$ we define the 1D Mellin transform of $I:\mathbb{R}_+\rightarrow \mathbb{C}$, where $\mathbb{R}_+:=[0,\infty)$,   by
\begin{equation}\label{eq:MT-intro}
\widetilde I(\nu):=\inty{0}{\infty}{I(\wp)\, \wp^{-\nu}}{\wp}~,\quad  \breve a< \Re\nu< \breve b~.	
\end{equation}
By use of~\eqref{eq:2-side-LT-INV}, the inverse Mellin transform of $\widetilde I$  gives
\begin{equation}\label{eq:MT-INV-intro}
I(\wp)=\frac{1}{2\pi\im}\int\limits_{\breve \gamma-\im \infty}^{\breve\gamma+\im \infty}\wp^{\nu-1}\,\widetilde I(\nu)\ \db\nu~,\quad \breve a< \breve\gamma< \breve b~.	
\end{equation}

Without further ado, the $n$-dimensional Mellin transform of $I: (\mathbb{R}_+)^n \rightarrow \mathbb{C}$ is defined from the $n$-dimensional Laplace transform of $g:\mathbb{R}^n \rightarrow \mathbb{C}$ via the mapping $t\mapsto \wp=(\wp_1,\,\ldots,\,\wp_n):=(e^{t_1},\,\ldots,\,e^{t_n})$. We have the pair
\begin{equation}\label{eq:MT-multi-intro}
\widetilde I(\nu):=\prod_{j=1}^n\left(\int_0^\infty \db \wp_j \ \wp_j^{-\nu_j}  \right)	I(\wp)~,\quad (\Re\nu_1,\,\ldots,\,\Re\nu_n)\in \mathbb{D}~;
\end{equation}
\begin{equation}\label{eq:MT-multi-INV-intro}
I(\wp)=\prod_{j=1}^n\left(\frac{1}{2\pi \im}
\int\limits_{\breve\gamma_j-\im\infty}^{\breve \gamma_j+\im \infty}  \db\nu_j \ \wp^{\nu_j-1}  \right)	\widetilde I(\nu)~,\quad (\breve\gamma_1,\,\ldots,\,\breve\gamma_n)\in \mathbb{D}~.
\end{equation}
Note that in principle $\mathbb{D}$ defines a polyhedron in $\mathbb{R}^n$.

Next, we heuristically discuss via an example how the 1D Mellin transform can be used for the extraction of asymptotic expansions~\cite{ChengWu-book}. Consider the pair $(I, \widetilde I)$ by~\eqref{eq:MT-intro} and~\eqref{eq:MT-INV-intro}. For some constant $C_0$, the formula
\begin{equation*}
 I(\wp) \sim C_0 \, \wp^{-1} (\ln\wp)^{\kappa}\quad \mbox{as}\ \wp\to +\infty 	
\end{equation*}
holds if and only if
\begin{equation*}
 \widetilde I(\nu) \sim C_0 \,\Gamma(1+\kappa)\, \nu^{-1-\kappa} \quad \mbox{as}\ \nu\to 0^+~. 	
\end{equation*}
Here, $\kappa>-1$ while $\nu\to \nu_{\diamond}^+$ means that the complex variable $\nu$ approaches $\nu_{\diamond}$ with $\Re\nu> \Re\nu_{\diamond}$, i.e., from the half-plane to the right of the line $\{\Re\nu=\Re\nu_{\diamond}\}$. If the singularity $\nu_{\diamond}$ of $\widetilde I(\nu)$ is shifted from $0$ to any point of the negative real axis then the asymptotic formula for $I(\wp)$ is multiplied by a negative power of $\wp$. In summary, the underlying idea is stated roughly as follows: Logarithmic terms in the asymptotic expansion of $I(\wp)$ for large $\wp$ correspond to algebraic singularities of $\widetilde I(\nu)$ lying to the \emph{left} of the analyticity strip.

Thus, the Mellin transform is appealing because a power law (in some prescribed limit) is plausibly easier to describe in comparison to a logarithmic behavior~\cite{ChengWu-book}. If  some integral representation is used for $I(\wp)$, the contributions of logarithmic terms as $\wp\to\infty$ may come from the whole region of integration. In contrast, if $\kappa\in\mathbb{N}$ then the singular point $\nu_{\diamond}$ is a pole of $\widetilde I(\nu)$, which can be studied with relative ease. This technique can be powerful for obtaining the full asymptotic expansion made of terms of the form $\wp^s (\ln\wp)^{\kappa}$ for $I(\wp)$ as $\wp\to +\infty$. This expansion can be constructed from all the contributions of singularities of $\widetilde I(\nu)$ by shift of the inversion path for $I(\wp)$ to the \emph{left} of the initial strip of analyticity.

The above argument can be extended to the study of the asymptotic behavior of $I(\wp)$ as $\wp\to \wp_{\diamond}^+$, say, $\wp_{\diamond}=0$ (for $0<\wp\ll 1$). The idea is to shift the inversion path for $I(\wp)$ to the \emph{right} of the original strip of analyticity of $\widetilde I(\nu)$, and pick the relevant contributions, e.g., residues from poles.

Now let us discuss how these considerations can be transferred to a higher dimension $n$, for functions $I: (\mathbb{R}_+)^n \rightarrow \mathbb{C}$ where $n\ge 2$. A plausible procedure is suggested by the iterated integrals in~\eqref{eq:MT-multi-INV-intro}: By making a \emph{particular choice} of the order of integrations, one may carry out each of the $n$ 1D inverse Mellin transforms successively via truncation of the corresponding expansion. There are at least two possible difficulties in this task.  First, one must remain consistent with the restriction $\breve\gamma\in \mathbb{D}$. This is achieved via the successive projections of the region $\mathbb{D}$ by means of multivariable calculus.

The second difficulty is that the asymptotic expansion for $I(\wp)$ may depend on the chosen order of the iterated integrals in~\eqref{eq:MT-multi-INV-intro}. This issue is expected: Asymptotic expansions can be divergent series; thus, rearrangements of their terms may alter the outcome.
Our `rule of thumb' is to carry out first the integration with respect to the dual variable, $\nu_{j_*}$ for some $j_*$, that corresponds to the largest parameter, $\wp_{j_*}$. For the SSH model, $\wp_{j_*}=\epsilon_3$. We carry out last the integration in the dual variable that corresponds to an unrestricted parameter. For the SSH model, this parameter is $\epsilon_2$.

\setcounter{equation}{0}

\section{On the generalized zeta function}
\label{app:zeta}
In this appendix, we discuss the generalized zeta function $\zeta(\vartheta, 1/2)$, which enters the result of Proposition~\ref{prop:exact_MT}. In particular, we show~\eqref{eq:gen-zeta-R}  regarding the connection of $\zeta(\vartheta, 1/2)$ to the Riemann zeta function, $\zeta(\vartheta)$.

We start with the standard definition of $\zeta(\vartheta,\varsigma)$, viz.~\cite{Bateman-I},
\begin{equation*}
	 \zeta(\vartheta,\varsigma):=\sum_{n=0}^\infty (\varsigma+n)^{-\vartheta}~,\qquad  \Re\vartheta >1~,\quad -\varsigma\notin \mathbb{N}=\{0,\,1,\,\ldots\, \}~.
\end{equation*}
First, $\zeta(\vartheta, 1/2)$ has a simple pole at $\vartheta=1$. This is deduced from~\cite{Bateman-I}
\begin{equation*}
	\lim_{\vartheta\to 1}\left\{\zeta(\vartheta, \varsigma)-\frac{1}{\vartheta -1}\right\}=-\psi(\varsigma)~;\quad \psi(\varsigma):=\frac{\db}{\db \varsigma}\Gamma(\varsigma)~,\ \Re \varsigma >0~.
\end{equation*}
For $\varsigma=1/2$, the right-hand side becomes $\gamma+ 2\ln 2$ where $\gamma=0.577215\ldots$ is Euler's constant. Note that $\vartheta=1$ is the only pole of $\zeta(\vartheta, \varsigma)$~\cite{Bateman-I}.

In addition, if $\Re\vartheta<0$ the function $\zeta(\vartheta, \varsigma)$ with $\varsigma=1/2$ has the same zeros as $\sin(\pi\vartheta/2)$. This can be seen from the Hurwitz formula~\cite{Bateman-I}, viz.,
\begin{equation*}
	\zeta(\vartheta,\varsigma)=2 (2\pi)^{\vartheta-1}\Gamma(1-\vartheta)\sum_{m=1}^\infty m^{\vartheta-1}\,\sin(2\pi m \varsigma+\pi\vartheta/2)~;\ \Re\vartheta <0,\ 0< \varsigma\le 1~.
\end{equation*}

The above properties suggest an intimate connection between the functions $\zeta(\vartheta, 1/2)$ and $\zeta(\vartheta)$. To show their relation, thus recovering~\eqref{eq:gen-zeta-R}, we use the definition of $\zeta(\vartheta, \varsigma)$ at $\varsigma=1/2$ to write
\begin{align*}
	\zeta(\vartheta, 1/2)&= \sum_{m=0}^\infty \left(m+\textstyle{\frac12}\right)^{-\vartheta}=2^{\vartheta} \sum_{m=0}^\infty (2m+1)^{-\vartheta}\notag\\
	&= 2^{\vartheta}\left\{\sum_{m=0}^\infty (1+m)^{-\vartheta}-\sum_{m=0}^\infty (2+2m)^{-\vartheta}   \right\}\notag\\
	&= 2^\vartheta (1-2^{-\vartheta})\sum_{m=0}^\infty (m+1)^{-\vartheta}= (2^\vartheta-1)\,\zeta(\vartheta)~,\qquad \Re\vartheta >1~.
\end{align*}
These steps yield~\eqref{eq:gen-zeta-R}, which is analytically continued to all 
$\vartheta\in \mathbb{C}$.  

\setcounter{equation}{0}

\section{Evaluation of certain hypergeometric series}
\label{app:hyperg}
In this appendix, we compute two cases of the hypergeometric function, ${}_2F_1$, in terms of elementary functions. The results are invoked in the proof of Proposition~\ref{prop:zeroT-I} (Section~\ref{subsec:zero-T}). Recall that the hypergeometric function ${}_2F_1(a, b; c; z)$ is defined by the series (if $c\neq -n$, $\forall\ n\in\mathbb{N}$)
\begin{equation}\label{eq:app:2F1-def}
	{}_2F_1(a, b; c; z):=\sum_{n=0}^\infty \frac{z^n}{n!}\,\frac{(a)_n\,(b)_n}{(c)_n}~,\quad |z|<1~;\ (a)_n:=\frac{\Gamma(a+n)}{\Gamma(a)}~.
\end{equation}

First, we show relation~\eqref{eq:2F1-log}, which is needed in the computation of $I^{(0)}$ and $I^{(1)}$ (Section~\ref{subsec:zero-T}). Note the identity~\cite{Bateman-I}
\begin{align*}
	{}_2F_1(a, b; a+b; 1-z)&=\frac{\Gamma(a+b)}{\Gamma(a)\,\Gamma(b)}\sum_{n=0}^\infty\frac{z^n}{n!}\frac{(a)_n (b)_n}{(1)_n}\left\{2\psi(1+n)-\psi(a+n) \right. \notag\\
	& \left. \qquad -\psi(b+n)-\ln(z) \right\}
\end{align*}
which, for $a=3/2$ and $b=1$, we used in order to write the initial series for $I^{(0)}$ in terms of ${}_2F_1(\frac32, 1; \frac52, 1-\epsilon_1/\epsilon_2)$; cf.~\eqref{eq:I-leading}. Now consider a particular Gauss' linear relation among continguous hypergeometric functions~\cite{Bateman-I}, viz.,
\begin{align*}
	c(1-z)\{{}_2F_1(a,b;c;z)\}-c\{{{}_2F_1(a-1,b;c;z)}\}+(c-b)z \{{{}_2F_1(a,b; c+1; z)}\}=0,
\end{align*}
and set $a=3/2$, $b=1$ and $c=3/2$. Thus, we obtain
\begin{equation}\label{eq:app:continguous}
{}_2F_1\bigl({\textstyle\frac32}, 1; {\textstyle\frac52}; z\bigr)=\frac{3}{z}\left\{{}_2F_1\bigl({\textstyle\frac12}, 1; {\textstyle\frac32}; z\bigr)-(1-z)\, {}_2F_1\bigl({\textstyle\frac32}, 1; {\textstyle\frac32}; z\bigr)\right\}~.	
\end{equation}
The hypergeometric functions of the right-hand side can be computed by
\begin{equation}\label{eq:app:conting1}
{}_2F_1\bigl({\textstyle\frac32}, 1; {\textstyle\frac32}; z\bigr)=\sum_{n=0}^\infty \frac{z^n}{n!}\,(1)_n=\frac{1}{1-z}~,	
\end{equation}
\begin{equation}\label{eq:app:conting2}
{}_2F_1\bigl({\textstyle\frac12}, 1; {\textstyle\frac32}; w^2\bigr)=\frac{1}{2w}\,\ln\biggl(\frac{1+w}{1-w}\biggr)~.	
\end{equation}
In the last equation we must set $w^2=z$, in view of~\eqref{eq:app:continguous}. Then, the desired relation~\eqref{eq:2F1-log} is recovered from~\eqref{eq:app:continguous}--\eqref{eq:app:conting2}.

Next, let us show formula~\eqref{eq:I2-F}. To this end, we apply the identity~\cite{Bateman-I}
\begin{equation*}
	{}_2F_1\bigl(1+{\textstyle\frac{\eta}{2}}, 1-{\textstyle\frac{\eta}{2}}; {\textstyle\frac32}; (\sin w)^2\bigr)=\frac{2\sin(\eta w)}{\eta \sin (2w)}~.
\end{equation*}
In the limit $\eta\to 0$, with fixed $w$, this relation yields
\begin{equation}\label{eq:app:2F1-sinw}
	{}_2F_1\bigl(1, 1; {\textstyle\frac32}; (\sin w)^2\bigr)=\frac{2w}{\sin (2w)}=\frac{w}{(\sin w)\,(\cos w)}~.
\end{equation}
%
Now map $w\mapsto z$ with $z=(\sin w)^2$ which entails $w=\sin^{-1}(\sqrt{z})$, in a suitable branch of the function $\sin^{-1}(\sqrt{z})$. By this replacement, \eqref{eq:app:2F1-sinw} leads to~\eqref{eq:I2-F}.

\printbibliography  

\end{document}